\documentclass[a4paper,11pt]{article}
\pdfoutput=1 
\usepackage{amsmath, amssymb}
\DeclareUnicodeCharacter{2A7E}{\gtrsim}
\usepackage{jcappub} 
\usepackage[T1]{fontenc} 
\usepackage{multirow}
\usepackage{aas_macros}
\usepackage{orcidlink}
\usepackage{subcaption}
\usepackage{todonotes}
\usepackage{hyperref}
\usepackage{academicons}
\usepackage{xcolor}
\usepackage{soul}
\usepackage{booktabs}
\sethlcolor{yellow}
\definecolor{orcidlogocol}{HTML}{A6CE39}
\usepackage{caption}
\usepackage{silence}

\title{Multi-Messenger Cosmology: A Route to Accurate Inference of Dark Energy Beyond CPL Parametrization from XG Detectors}

\author{Samsuzzaman Afroz \orcidlink{0009-0004-4459-2981},}
\author{Suvodip Mukherjee \orcidlink{0000-0002-3373-5236}}

\affiliation{Department of Astronomy and Astrophysics, Tata Institute of Fundamental Research, 1, Homi Bhabha Road, Mumbai 400005, India}

\emailAdd{samsuzzaman.afroz@tifr.res.in}
\emailAdd{suvodip@tifr.res.in}

\abstract{One of the central challenges in modern cosmology is understanding the nature of dark energy and its evolution throughout the history of the Universe. Dark energy is commonly modeled as a perfect fluid with a time-varying equation-of-state parameter, \( w(z) \), often modeled under CPL parametrization using two parameters \( w_0 \) and \( w_a \). In this study, we explore both parametric and non-parametric methods to reconstruct the dark energy Equation of State (EoS) using Gravitational Wave (GW) sources, with and without electromagnetic (EM) counterparts called bright sirens and dark sirens respectively. In the parametric approach, we extend the widely used \( w_0 \)-\( w_a \) model by introducing an additional term, \( w_b \), to better capture the evolving dynamics of dark energy up to high redshift which is accessible from GW sources. This extension provides increased flexibility in modeling the EoS and enables a more detailed investigation of dark energy’s evolution. Our analysis indicates that, with five years of observation time and a 75\% duty cycle using Cosmic Explorer and the Einstein Telescope, it will be possible to measure the dark energy EoS with remarkable precision, better than any other cosmological probe in the coming years, from bright standard sirens using a multi-messenger approach. These findings highlight the potential of GW observations in synergy with EM telescopes to offer valuable insights into the nature of dark energy, overcoming the current limitations in cosmological measurements. }

\begin{document}
\maketitle
\flushbottom
\section{Introduction}
\label{sec:Introduction}

The nature of dark energy, a mysterious component responsible for approximately 70\% of the total energy density of the universe, remains one of the most profound puzzles in modern cosmology \citep{Huterer:2017buf,SupernovaCosmologyProject:1996grv,SupernovaSearchTeam:1998bnz}. Its existence was first inferred from observations of the accelerated expansion of the universe \citep{SupernovaSearchTeam:1998fmf,SupernovaCosmologyProject:1998vns}, and since then, understanding its properties has become a central focus of cosmological research. The simplest candidate for driving this acceleration is the Cosmological Constant, \(\Lambda\), as proposed in the \(\Lambda\)CDM model \citep{Carroll:2000fy,SupernovaSearchTeam:1998fmf,SupernovaCosmologyProject:1998vns,WMAP:2003elm,SDSS:2005xqv}. Despite its success in describing a wide range of cosmological observations, the \(\Lambda\)CDM model raises several theoretical questions that continue to motivate investigations into the nature of \(\Lambda\). These include questions about the value of this constant and the specific moment in the universe's history when it began to dominate over other components \citep{Martin:2012bt,Bernardo:2022cck,Sahni:2002kh,Weinberg:2000yb,Capozziello:2018jya}. 

Recent observational precision has begun to reveal tensions in the determination of \(\Lambda\)CDM parameters, especially when measurements are obtained from different cosmological probes \citep{Abdalla:2022yfr,SDSS:2003eyi}. One of the most striking discrepancies is seen in the values of the Hubble constant, \(H_0\), which differ between local measurements \citep{2022ApJ...934L...7R} and those inferred from Cosmic Microwave Background (CMB) observations \citep{Planck:2018vyg}. While the former directly measure the current rate of expansion, the latter rely on the assumption of the \(\Lambda\)CDM model to extrapolate high-redshift data to the present. Numerous alternative models to the \(\Lambda\)CDM framework have been tested against observations, often leading to the exclusion of specific theories. More commonly, however, these alternatives only provide constraints on their parameter spaces that are consistent with the \(\Lambda\)CDM model. Given the vast number of potential models and the challenges in testing all of them, a model-independent approach is increasingly desirable. The first efforts toward such an approach began nearly two decades ago, following the discovery of the late-time acceleration of the universe.

The equation of state (EoS) parameter \( w(z) \), which relates the pressure and density of dark energy, is commonly used to characterize its behavior \citep{Frampton:2002tu,Bastero-Gil:2002plx}. One of the most widely adopted parametrizations for \( w(z) \) is the Chevallier-Polarski-Linder (CPL) model, which assumes a simple linear evolution of the EoS with redshift, defined as \( w(z) = w_0 + w_a \frac{z}{1+z} \) \citep{Chevallier:2000qy,Linder:2002et,dePutter:2008wt}. While the CPL parametrization has been useful for describing the time-dependent nature of dark energy, it remains limited in its ability to capture the full complexity of dark energy’s evolution throughout cosmic history \cite{Shlivko:2024llw, PhysRevD.108.103519}. This model provides a two-parameter description of the EoS for dark energy, a component that dominates the expansion at late times and drives cosmic acceleration. This component can be an additional fluid contributing to the energy-momentum tensor, or it may result from modifications of gravity. Although the CPL parameterization is widely used to constrain the time evolution of dark energy and to assess the potential of future experiments, it still relies on specific assumptions regarding the time evolution of \( w(z) \). An alternative approach is the "non-parametric" reconstruction of \( w(z)\), where its value is constrained at different redshifts \citep{2010PhRvD..82j3502H,2012PhRvD..85l3530S,2012JCAP...06..036S,Keeley:2019esp}.

\begin{figure}[ht]
    \centering
    \includegraphics[height=11.0cm, width=12cm]{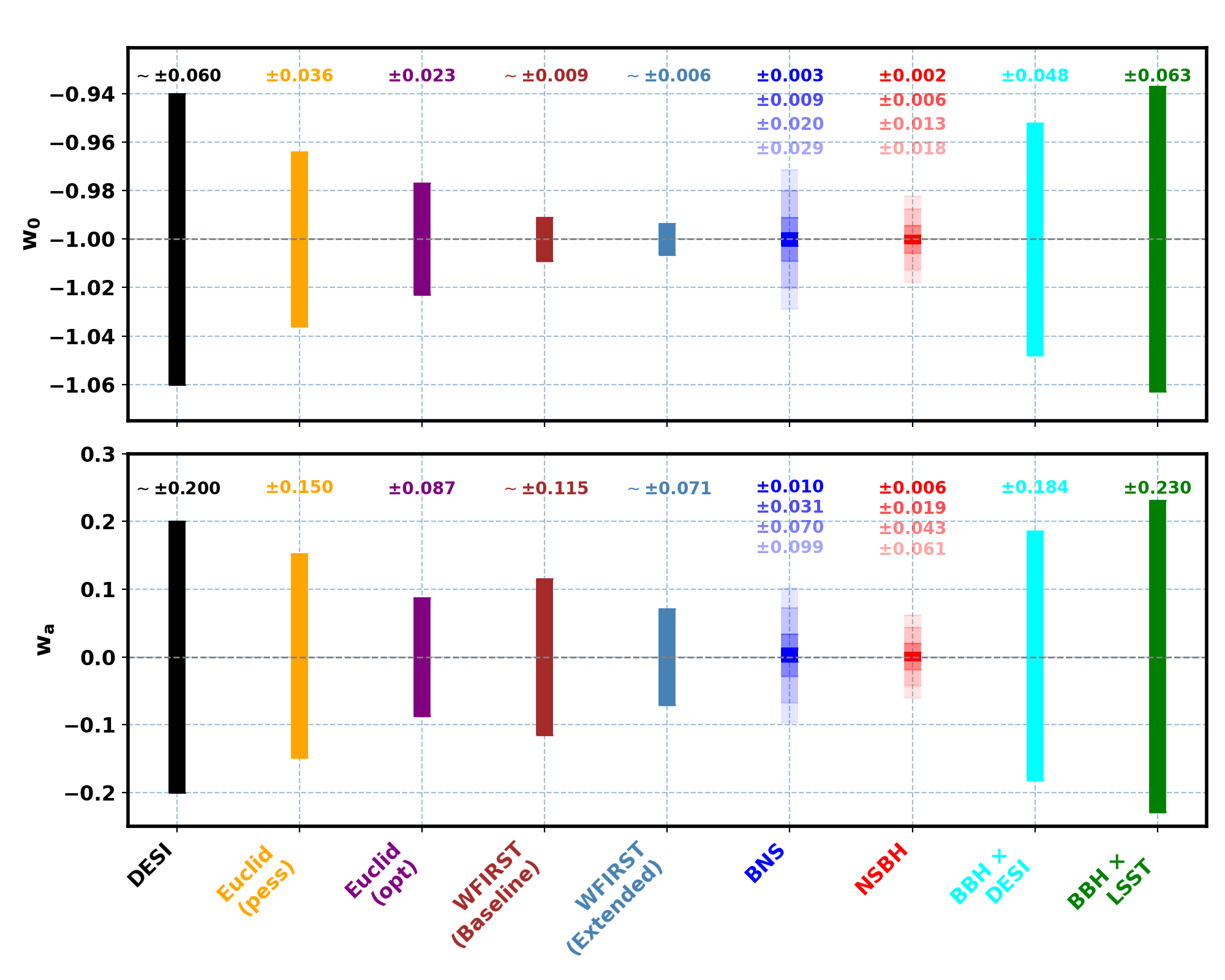}
    \caption{This figure illustrates the projected constraints on the dark energy equation of state parameters, $w_0$ and $w_a$, from various cosmological surveys, including Euclid, DESI, WFIRST, and the upcoming GW observatories, Cosmic Explorer and Einstein Telescope. For the GW scenarios, BNS, NSBH, and BBH systems are considered, assuming fixed values of $H_0 = 67.4 \, \mathrm{km \, s^{-1} \, Mpc^{-1}}$ and $w_b = 0$. To estimate the redshift of BBH events, cross-correlation is performed using galaxy catalogs from DESI (BBH$\times$DESI) and LSST (BBH$\times$LSST). The error bars represent $1\sigma$ uncertainties, and the surveys are color-coded, with each color corresponding to a specific observational program. Horizontal and vertical reference lines are included to highlight deviations from the standard $\Lambda$CDM cosmological model ($w_0 = -1$, $w_a = 0$). For BNS and NSBH scenarios, the constraints depend upon the fraction of EM counterpart detections associated with these events. This figure shows four different cases of EM detections: the faintest (lightest) corresponds to 1\% detection, followed by 10\% detection, 50\% detection, and the darkest represents 100\% detection of EM counterparts. These variations illustrate how increasing EM detection rates lead to better constraints on dark energy EoS parameters. The error bars for WFIRST and DESI were visually extracted from \href{https://ui.adsabs.harvard.edu/abs/2013arXiv1305.5425S/abstract}{figure 5} of \cite{2013arXiv1305.5425S} and \href{https://arxiv.org/abs/1611.00036}{figure 2.11} of \cite{DESI:2016fyo}, respectively, while the constraints for Euclid were taken from \href{https://arxiv.org/abs/2105.09746}{Table 3} of \cite{Euclid:2021cfn}, corresponding to the $\Lambda$CDM fiducial model. The details on the results are mentioned in section \ref{sec:DEosReconstructon}.}
    \label{fig:SurveyComp}
\end{figure}

In recent years, Gravitational Wave (GW) observations \citep{LIGOScientific:2016aoc,LIGOScientific:2017vwq,LIGOScientific:2020iuh,LIGOScientific:2020zkf,LIGOScientific:2019zcs,LIGOScientific:2021aug,LIGOScientific:2017adf} have emerged as a novel and powerful tool for cosmology, particularly for studying dark energy. GWs offer an independent means of measuring the luminosity distance which does not need to be calibrated. As a result, mapping the expansion history accross cosmic redshifts can be done without using a distance ladder  such as for the traditional methods, such as Type Ia supernovae (SNe Ia). By measuring the distances to high-redshift sources within the framework of General Relativity (GR), GWs provide valuable insights. However, these inferred distances can also vary due to the effects of modified gravity theories \citep{Belgacem:2018lbp,Mukherjee:2020mha,Afroz:2024joi,Afroz:2024oui,Afroz:2023ndy}. As GW astronomy progresses, the combination of GW measurements with other cosmological data offers the potential to provide a more refined reconstruction of the dark energy EoS. 

\begin{figure}[ht]
    \centering
    \includegraphics[height=8.0cm, width=16cm]{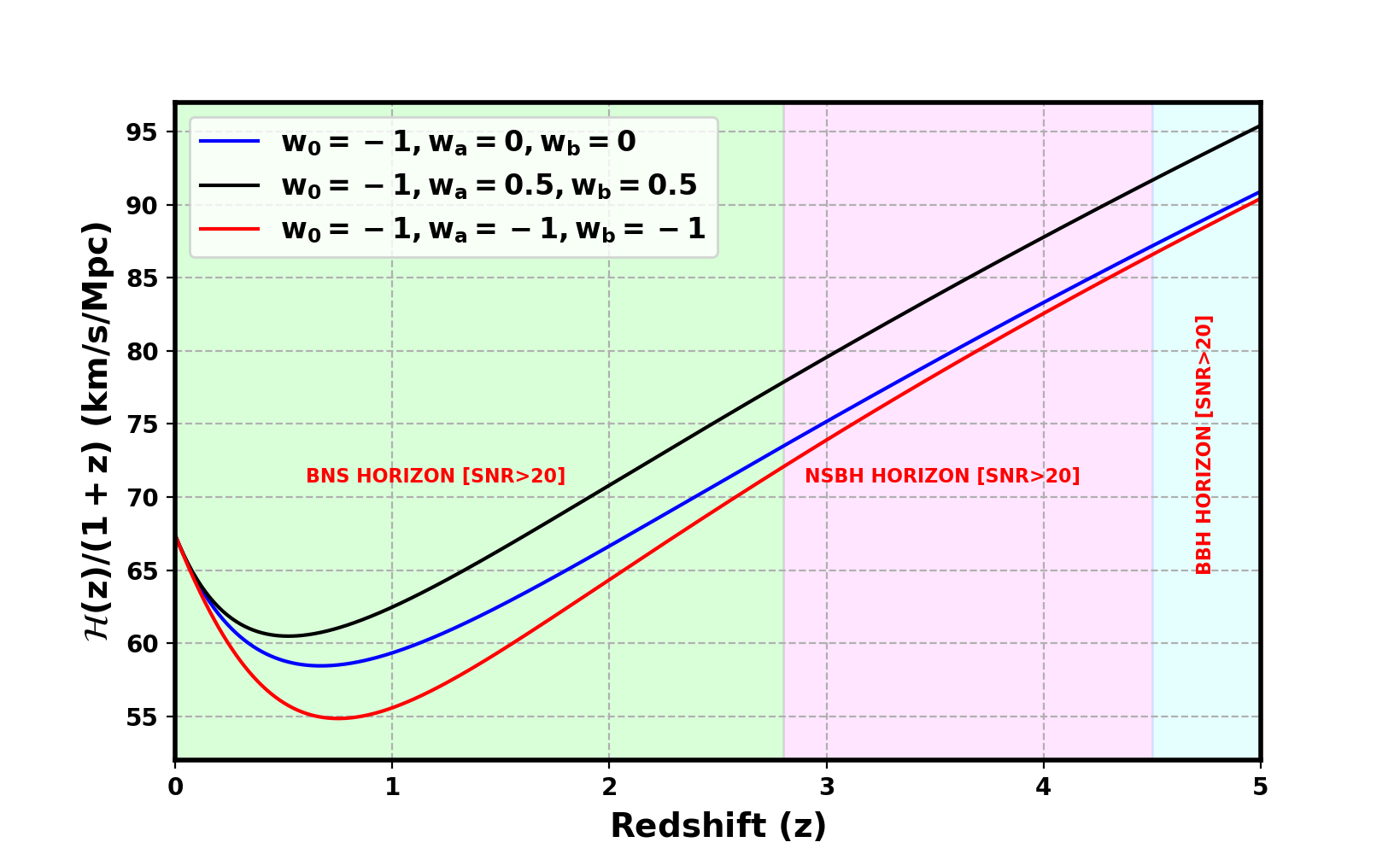}
    \caption{The plot shows the Hubble parameter, more specifically $\mathcal{H}(z)/(1+z)$, as a function of redshift $z$ under different dark energy equation of state parametrizations: $w_0 = -1$, $w_a = 0, w_b = 0$ (blue), $w_0 = -1$, $w_a = 0.5$, $w_b = 0.5$ (black), and $w_0 = -1$, $w_a = -1$, $w_b = -1$ (red). The shaded regions represent the detection horizons for various GW event categories assuming a signal-to-noise ratio threshold of $>20$. These regions illustrate how various types of GW sources contribute to constraining cosmological models and understanding the evolution of GW events over cosmic time.}
    \label{fig:HorizonPlot}
\end{figure}

In figure \ref{fig:SurveyComp}, we show the key findings of our analysis and its comparison with other upcoming cosmic probes to dark energy from the future GW detectors, such as LISA \citep{LISACosmologyWorkingGroup:2022jok}, the Cosmic Explorer (CE) \citep{Reitze:2019iox}, and the Einstein Telescope (ET) \citep{Branchesi:2023mws}, which will be capable of detecting GW events at significantly higher redshifts as shown in figure \ref{fig:HorizonPlot}. The details on this plot is discussed in the section \ref{sec:DEosReconstructon} where we discuss the results. This extended range enables us to investigate the Universe's expansion history and the dynamics of dark energy in regimes that are difficult to probe with traditional methods. In particular, the standard siren approach based on the luminosity distance–redshift relation \cite{Arabsalmani:2013bj,Mitra:2020vzq,Vitale:2018wlg,LISACosmologyWorkingGroup:2019mwx,ET:2019dnz,Mukherjee:2020hyn, Diaz:2021pem,Cozzumbo:2024vxw,Jin:2021pcv,Jin:2022qnj,Jin:2020hmc} and multi-band gravitational-wave observations, which track the Universe's aging \cite{Mukherjee:2024inw}, provide complementary avenues for studying dark energy. This capability allows us to move beyond the standard CPL parametrization framework and explore more general, dynamical models of dark energy. Except at redshifts close to zero, higher-order terms in the dark energy EoS can become significant, offering a richer and more accurate description of its behavior over cosmic time.

To account for potential deviations from the simple linear evolution assumed in the traditional CPL parametrization , we extend the model by introducing an additional parameter, \(w_b\), thereby constructing a three-parameter framework for the dark energy EoS \(w(z)\) \citep{AlbertoVazquez:2012ofj, Linder:2005ne, Clarkson:2007bc, Ratra:1987rm, Caldwell:1997ii, Caldwell:1999ew, Copeland:2006wr}. This extended model offers the flexibility to capture potential curvature or non-linear evolution in \(w(z)\) that may become detectable as observational data extend to higher redshifts and achieve percent-level accuracy. Importantly, this extension serves as a diagnostic tool: if \(w_b\) is statistically consistent with zero, it reinforces the robustness of the standard two-parameter CPL description; conversely, a statistically significant deviation would indicate the presence of richer dark energy dynamics. By combining GW observations with other cosmological data, we aim to refine the reconstruction of the dark energy EoS and study its evolution over a wide range of redshifts. Using this extended framework, we demonstrate that the parameters in both the standard and extended CPL models can be effectively constrained by future ground-based GW observatories. Moreover, our approach aids in breaking parameter degeneracies when combining multiple datasets and complements principal component analyses, which generally indicate that only a few modes are tightly constrained. Furthermore, we perform a model-independent reconstruction of the dark energy EoS and the Hubble parameter. These findings show a transformation in our understanding of dark energy from next-generation GW observatories like CE and ET in synergy with EM telescopes.

In this analysis, we incorporate both bright and dark sirens to study the dark energy EoS, adopting a generalized approach that does not rely on a specific type of source. Bright sirens, which have identifiable electromagnetic (EM) counterparts, enable direct redshift measurements. These EM counterparts can be observed through coordinated follow-up observations using a variety of telescopes and missions, such as the Hubble Space Telescope \citep{Scoville:2006vr},  Fermi \citep{Bissaldi:2008df}, Swift \citep{SWIFT:2005ngz}, LSST \citep{2009arXiv0912.0201L}, and Zwicky Transient Facility \citep{2014htu..conf...27B}. These missions can detect the light signals associated with GW events, providing crucial redshift information. Coordinated follow-up efforts with these observatories can confirm the location and properties of GW events, improving redshift measurements and enabling a deeper understanding of cosmological parameters.
In contrast, dark sirens, lacking EM counterparts, require statistical methods such as cross-correlation with galaxy surveys to infer redshifts for measuring dark energy \cite{Mukherjee:2018ebj,Oguri:2016dgk, Mukherjee:2019wcg, Diaz:2021pem,balaudo2023prospects}. To explore the impact of different types of galaxy surveys, we utilized two major surveys: the Dark Energy Spectroscopic Instrument (DESI) \citep{DESI:2023ytc}, a spectroscopic survey, and the Vera C. Rubin Observatory’s Legacy Survey of Space and Time (LSST) \citep{2009arXiv0912.0201L}, a photometric survey. This comparison allowed us to examine how the survey type (spectroscopic versus photometric) influences the results. By leveraging these complementary methods and data sources, we demonstrate the unique capability of GW observations to probe the universe's expansion history and provide deeper insights into the nature of dark energy.

This paper is organized as follows: section \ref{sec:Framework} introduces the framework for reconstructing the dark energy EoS using both parametric and non-parametric approaches. Section \ref{sec:MockSamples} focuses on the astrophysical population of GW sources used in this analysis. In section \ref{sec:RedInf}, we detail the methodology for inferring the redshifts of these GW sources. Section \ref{sec:HzReconstruction} presents the reconstruction of the Hubble parameter from GW observations. In section \ref{sec:DEosReconstructon}, we demonstrate the reconstruction of the dark energy EoS using both parametric and non-parametric techniques. Finally, section \ref{sec:Summery} summarizes our key findings, highlights the strengths of the proposed methodology, and discusses future directions.

\section{Framework for Dark Energy Equation of State}
\label{sec:Framework}

In standard cosmology, the expansion of the universe is governed by the Friedmann equations, which describe the evolution of the Hubble parameter \( \mathcal{H}(z) \). A fundamental equation that relates the redshift \( z \) of a source to its luminosity distance is the luminosity distance redshift relation:
\begin{equation}
    D_L(z) = (1+z) \int_0^z \frac{c}{\mathcal{H}(z')} \, \mathrm{d}z',
    \label{eq:DistRedshift}
\end{equation}
where \( c \) is the speed of light, \( \mathcal{H}(z) \) is the Hubble parameter at redshift \( z \), and \( D_L(z) \) is the luminosity distance to the source at redshift \( z \).

The evolution of the Hubble parameter is influenced by the densities of matter, radiation, and dark energy, as well as the equation of state of dark energy \( w(z) \). Assuming a spatially flat universe (i.e., zero curvature, so that \( \Omega_\text{DE} = 1 - \Omega_m - \Omega_r \)), the behavior of \( \mathcal{H}(z) \) is described by the following equation:
\begin{equation}\label{eq:hzde}
\mathcal{H}^2(z) = H_0^2 \left[ \Omega_m (1+z)^3 + \Omega_r (1+z)^4 + \Omega_\text{DE} \exp\left(3 \int_0^z \frac{1+w(z')}{1+z'} \, dz'\right) \right],
\end{equation}
where \( H_0 \) is the Hubble constant, representing the current expansion rate of the universe, \( \Omega_m \) and \( \Omega_r \) are the present-day normalized densities of matter and radiation, respectively, and \( \Omega_\text{DE} \) is the normalized density of dark energy. The exponential term accounts for the contribution of dark energy, with \( w(z) \) being its EoS parameter as a function of redshift. The EoS relates the pressure \( p(z) \) and energy density \( \rho(z) \) of dark energy through the relation:
\begin{equation}
    p(z) = w(z) \rho(z),
\end{equation}
where \( w(z) \) serves as a crucial parameter in understanding the nature and evolution of dark energy as the universe expands. Substituting the expression for \( \mathcal{H}(z) \) into the luminosity distance formula, we obtain:

\begin{equation}
    D_L(z) = (1+z) \int_0^z \frac{c \, dz'}{H_0 \sqrt{\Omega_m (1+z')^3 + \Omega_r (1+z')^4 + \Omega_\text{DE} \exp\left(3 \int_0^{z'} \frac{1+w(z'')}{1+z''} \, dz''\right)}}.
\end{equation}
This equation connects the dark energy EoS \( w(z) \) to the observable quantities, namely the redshift \( z \) and the luminosity distance \( D_L(z) \). By analyzing these relationships, we can infer information about the expansion history of the universe and the properties of dark energy. The standard approach to modeling the evolution of dark energy EoS is through the CPL parametrization, which expresses the EoS as a function of redshift:
\begin{equation}
    w(z) = w_0 + w_a \frac{z}{1+z},
\end{equation}
where \( w_0 \) represents the current value of the EoS, and \( w_a \) governs its redshift dependence. 

The CPL model, while widely recognized for its simplicity and adaptability in fitting observational data, particularly in the low-redshift regime, has significant limitations in capturing more complex behaviors at higher redshifts or deviations from its functional form. The interplay between dark energy dynamics and the expansion history, especially in thawing or freezing quintessence scenarios, has underscored the need for more robust modeling approaches \citep{Scherrer:2007pu,Dutta:2008qn,Chiba:2009sj}. Furthermore, the intricate dynamics of scalar field potentials are often oversimplified by parametric models like CPL, potentially leading to a misrepresentation of the true nature of dark energy \citep{Wetterich:1994bg,Bedroya:2019snp,Andrei:2022rhi,Zhang:2005eg}. These limitations highlight the importance of pursuing alternative, physically motivated models or non-parametric methods that can better account for the diverse behaviors of dark energy across cosmic time, particularly in light of the high-precision data expected from upcoming surveys\citep{Shlivko:2024llw,Sahni:2006pa}.

To address the limitations of the standard CPL model, we consider two cases, (i) model-independent reconstruction of dark energy EoS that avoids assuming any specific functional form. Instead, this approach reconstructs \( w(z) \) directly from observed luminosity distances and redshifts. The primary advantage of this framework is its flexibility, as it allows for inference of the evolution of dark energy without relying on predefined models or parametrization; and (ii) extend the CPL parametrization by introducing an additional parameter \( w_b \), which accounts for quadratic evolution of the dark energy EoS with redshift, given by:
\begin{equation}
    w(z) = w_0 + w_a \left(\frac{z}{1+z}\right) + w_b \left(\frac{z}{1+z}\right)^2,
    \label{eq:DEParametrization}
\end{equation}
where the third term \( w_b \left(\frac{z}{1+z}\right)^2 \) provides added flexibility in the description of dark energy evolution. This extended parametrization allows for a more accurate representation of dark energy dynamics at  redshifts beyond close to zero and offers a better fit to observational data than $w_0-w_a$. Application of this parametrization on GW data will allow to study any deviation from the CPL model and will be able to shed light on scenarios that are not well captured otherwise \cite{Shlivko:2024llw, PhysRevD.108.103519}. While higher-order terms could further refine the model, we limit this analysis to second-order terms for this study.

\section{Gravitational Waves  mock samples}
\label{sec:MockSamples}


The study of GW is set to be revolutionized by next-generation ground-based detectors such as the Cosmic Explorer \cite{Evans:2021gyd} and the Einstein Telescope \cite{Branchesi:2023mws}. These detectors are designed to vastly expand the observable volume of the universe and probe the low-frequency regime critical for precision cosmology. The CE, planned for construction in the United States, will feature arm lengths of 20 or 40 kilometers, significantly improving sensitivity to GW events from high-redshift sources and facilitating precise measurements of source parameters. Similarly, the ET, a European project with a unique triangular configuration and 10-kilometer-long underground arms, will minimize seismic noise and enable observations in the frequency range from a few Hz to several kHz. Together, these facilities will not only enhance our understanding of compact object mergers but also provide unprecedented opportunities for reconstructing the properties of dark energy through standard sirens.


To reconstruct the dark energy EoS, it is crucial to rely on a well-modeled and comprehensive approach to GW mock samples. The ability to accurately perform this reconstruction depends on a robust understanding of the astrophysical population of GW sources. The number of detectable GW events will depend on key factors such as merger rates, redshift distributions, and the mass populations of these sources. Next-generation GW observatories like the CE and ET will offer enhanced sensitivity, allowing them to detect binary mergers over an extended redshift range. This makes these observatories particularly powerful for conducting cosmological studies. In this analysis, we perform the dark energy EoS reconstruction using both bright and dark sirens. Bright sirens correspond to GW events with EM counterparts, such as BNS and NSBH mergers, where the redshift can be directly measured from their EM counterpart observations with spectroscopic precision.

In contrast, dark sirens are primarily BBH mergers without EM counterparts, where redshift information must be statistically inferred. A precise understanding of the population properties of GW sources is essential, as these properties directly affect the inferred redshift range and the accuracy of the reconstructed Hubble parameter $\mathcal{H}(z)$ and the dark energy EoS, $w(z)$. Our analysis incorporates detailed astrophysical models that capture GW source distributions, merger rates, and potential selection effects. These models are specifically tailored to the anticipated capabilities of next-generation GW observatories like CE and ET. In the following sections, we will systematically discuss the key ingredients of these GW mock samples, their statistical properties, and how they influence the reconstruction analysis. This comprehensive approach ensures the robustness of our cosmological analysis while accounting for the physical and observational factors that underpin GW detections.

\texttt{Mass model:} The mass population model for gravitational wave sources, including BBH, NSBH, and BNS systems, is informed by recent findings from the third catalog of GW sources published by the LVK collaboration \cite{abbott2023population, talbot2018measuring, abbott2019binary}. For black hole masses, we adopt the \textbf{Power Law + Gaussian Peak} model, which incorporates a power-law distribution to represent the primary mass distribution and a Gaussian component to account for the concentrated occurrences of black hole masses within the range \(5\,M_\odot\) to \(50\,M_\odot\). This model also includes additional smoothing to ensure a realistic and continuous distribution, consistent with current observational data. In particular, the mass distribution is given by \citep{KAGRA:2021duu}

\begin{equation}
p_{\rm BH}(m) = s(m) \left[ (1-\lambda_m)\frac{m^{-\alpha}}{N} + \lambda_m\, G(m) \right],
\end{equation}
where
\begin{equation}
N = \int_{m_{\min}}^{m_{\max}} m^{-\alpha}\, dm,
\end{equation}
\begin{equation}
G(m) = \frac{1}{\sqrt{2\pi\sigma_m^2}} \exp\left(-\frac{(m-\mu_m)^2}{2\sigma_m^2}\right),
\end{equation}
and the smoothing function \( s(m) \) is defined as
\begin{equation}
s(m) =
\begin{cases}
0, & m < m_{\min}, \\[1ex]
\displaystyle \frac{1}{1+\exp\left[\frac{\delta_m}{m-m_{\min}} + \frac{\delta_m}{(m-m_{\min})-\delta_m}\right]}, & m_{\min} \le m < m_{\min}+\delta_m, \\[1ex]
1, & m \ge m_{\min}+\delta_m.
\end{cases}
\end{equation}
The parameter values are set to \(\alpha = 2.3\), \(m_{\min} = 5\,M_\odot\), \(m_{\max} = 50\,M_\odot\), \(\lambda_m = 0.1\), \(\mu_m = 40\,M_\odot\), \(\sigma_m = 2\,M_\odot\), and \(\delta_m = 2\,M_\odot\). For neutron star masses, we assume a uniform mass distribution ranging from \( \rm{1M_{\odot}} \) to \( \rm{2M_{\odot}} \), reflecting the relatively narrow range of observed neutron star masses in binary systems \citep{KAGRA:2021duu}. By combining these distributions, the model provides a comprehensive framework for characterizing the mass populations of BBH, NSBH, and BNS systems in our analysis.

\texttt{Merger rate:}  In this study, we adopt a delay time model to describe the merger rates of various compact binary systems, including BNS, NSBH, and BBHs \cite{o2010binary, dominik2015double}. The delay time model characterizes the merger rate through a delay time distribution, which represents the time between the formation of the progenitor system and the eventual merger. This delay time depends on the evolutionary processes of the progenitor stars and the dynamics of the binary system, and it is important to note that the delay time is not uniform across all systems. Different types of compact binary systems can exhibit distinct delay time distributions and merger rates. For instance, BNS and NSBH systems may experience different evolutionary pathways and delay time distributions compared to BBH due to variations in stellar evolution processes, supernova mechanisms, and environmental factors. These differences lead to variations in their respective merger rates and the range of observed delay times. Therefore, accurately modeling these distributions and rates is essential for understanding the observed population of GW events and their underlying astrophysical properties. The delay time distribution functions used in this study capture these variations, providing a comprehensive statistical description of these processes across all compact binary systems. The distribution function accounts for the variations in the delay time and is defined as follows

\begin{equation}
    \mathrm{p_t(t_d|t_d^{min},t_d^{max},d) \propto 
    \begin{cases}
    (t_d)^{-d} & \text{, for }  t_d^{min}<t_d<t_d^{max}, \\
    0 & \text{otherwise}.    
    \end{cases}}
\end{equation}
The delay time is given by $\rm{t_d=t_m-t_f}$, where $\rm{t_m}$ and $\rm{t_f}$ are the lookback times of merger and formation respectively \cite{karathanasis2023binary}. So the merger rate at redshift $\rm{z}$ can be defined as
\begin{equation}
    \mathrm{R_{TD}(z)=R_0\frac{\int_z^{\infty}p_t(t_d|t_d^{min},t_d^{max},d)R_{SFR}(z_f)\frac{dt}{dz_f}dz_f}{\int_0^{\infty}p_t(t_d|t_d^{min},t_d^{max},d)R_{SFR}(z_f)\frac{dt}{dz_f}dz_f}}. 
\end{equation}
The parameter \( \rm{R_0} \) represents the local merger rate, which quantifies the frequency of mergers occurring at a redshift of \( \rm{z = 0} \). According to the study in \cite{abbott2023population}, the estimated \( \rm{R_0} \) values for BBH mergers range from 17.9 \( \rm{Gpc^{-3}\,yr^{-1}} \) to 44 \( \rm{Gpc^{-3}\,yr^{-1}} \). For BNS systems, \( \rm{R_0} \) varies widely, between 10 \( \rm{Gpc^{-3}\,yr^{-1}} \) and 1700 \( \rm{Gpc^{-3}\,yr^{-1}} \). In the case of neutron star–black hole (NSBH) systems, the \( \rm{R_0} \) values are estimated to lie between 7.8 \( \rm{Gpc^{-3}\,yr^{-1}} \) and 140 \( \rm{Gpc^{-3}\,yr^{-1}} \). For the purposes of our study, we adopt a standard local merger rate of \( \rm{R_0} = 20\,\rm{Gpc^{-3}\,yr^{-1}} \) uniformly for BNS, NSBH, and BBH systems. The numerator of the expression involves the integration over redshift $\rm{z_f}$ from $\rm{z}$ to infinity, where $\rm{p_t(t_d|t_d^{min},t_d^{max},d)}$ is the delay time distribution, $\rm{R_{SFR}(z_f)}$ is the star formation rate, and $\rm{\frac{dt}{dz_f}}$ is the Jacobian of the transformation. The star formation rate (SFR) denoted by $\rm{R_{SFR}(z)}$ is determined using the Madau Dickinson model \cite{madau2014cosmic}.

\begin{figure}[ht]
    \centering
    \includegraphics[height=8.0cm, width=16cm]{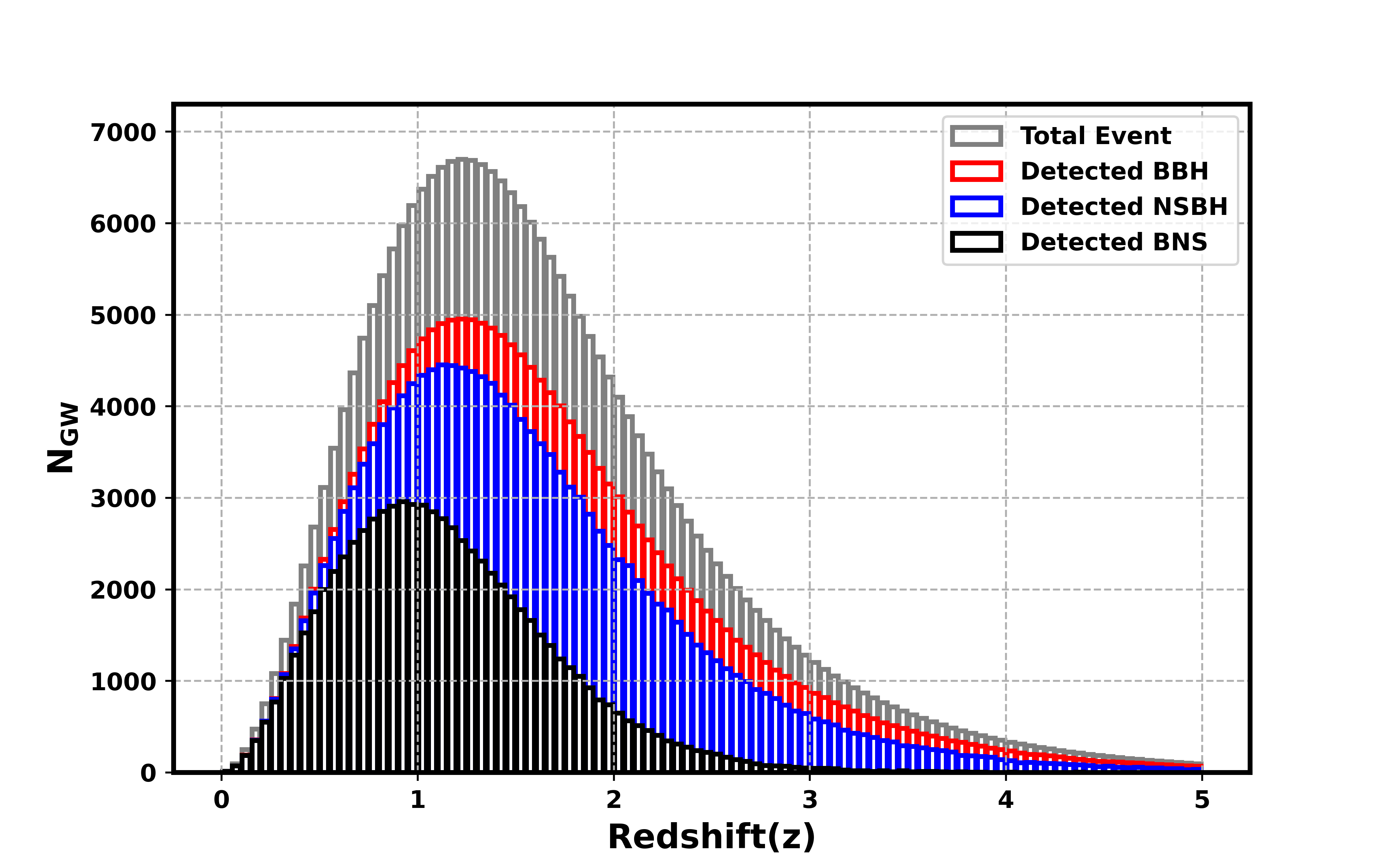}
    \caption{The figure illustrates the total number of simulated merger events (in gray) alongside the detectable events for BBH (in red), BNS (in black), and NSBH (in blue) using the CE+ET detector network. The analysis assumes a five-year observation period with a 75\% duty cycle. The total number of simulated merger events is 244,832, of which 180,221 BBH, 152,054 NSBH, and 75,832 BNS events are detected.}
    \label{fig:EventCount}
\end{figure}

The total number of compact binary coalescing events per unit redshift is estimated as 
\begin{equation}
\mathrm{\frac{dN_{GW}}{dz} = \frac{R_{\rm TD}(z)}{1+z} \frac{dV_c}{dz}(z) T_{obs}},
\label{eq:totno}
\end{equation}
where $T_{obs}$ indicates the total observation time, $\rm{\frac{dV_c}{dz}}$ corresponds to the differential comoving volume element, and $\rm{R(z)}$ denotes the merger rate \cite{karathanasis2022gwsim}. We consider the delay time merger rate with a specific minimum delay time $\rm{t_d =500\,\text{Myrs}}$ and a power-law exponent of $\rm{d=1}$. To determine which events are detectable, the calculation of the matched filtering signal-to-noise ratio (SNR) plays a crucial role. The SNR serves as a measure of the strength of the GW signal relative to the background noise. Only those events with a matched filtering SNR greater than or equal to a predetermined threshold SNR ($\rm{\rho_{\rm TH}}$) can be reliably detected. \cite{maggiore2007gravitational}. For a GW signal emitted by a coalescing binary system, the optimized SNR, denoted as $\rm{\rho}$, is defined as follows 
\begin{equation}
\mathrm{\rho^2 \equiv 4\int_{f_{\text{min}}}^{f_{\text{max}}} df \frac{|h(f)|^2}{S_n(f)}},
\label{snr}
\end{equation}
here, $\rm{S_n(f)}$ represents the noise power spectral density of the detector\cite{sathyaprakash1991choice, cutler1994gravitational, balasubramanian1996gravitational}. The function $\rm{h(f)}$ corresponds to the GW strain in the restricted post-Newtonian approximation and is defined for plus ($\rm{+}$) and cross ($\rm{\times}$) polarization as \cite{ajith2008template} 

\begin{equation}
    \mathrm{h(f)_{\{+, \times\}}=\sqrt{\frac{5\eta}{24}}\frac{(GM_c)^{5/6}}{D_L\pi^{2/3}c^{3/2}}f^{-7/6}e^{\iota\Psi(f)}\mathcal{I}_{\{+, \times\}}}.
\end{equation}

In this expression, the symbol $\rm{\eta}$ represents the symmetric mass ratio. The term $\rm{M_c}$ signifies the chirp mass of the system. The variable $\rm{D_L}$ denotes the luminosity distance. The constant c represents the speed of light in a vacuum. $\rm{\mathcal{I}_{+}= (1+\cos^{2}i)/2}$ and $\rm{\mathcal{I}_{\times}= \cos i}$ depends on the inclination angle $\rm{i}$. Finally, $\rm{\Psi(f)}$ stands for the phase of the waveform. However, the signal detected by a GW detector $\rm{h_{det}}$ is a complex interplay of several variables, including the detection antenna functions ($\rm{F_{+}, F_{\times}}$), and can be expressed as 
\begin{equation}
    \mathrm{h_{det}=F_{+}h_{+}+F_{\times}h_{\times}},
\end{equation}
here $\rm{F_{+}}$ and $\rm{F_{\times}}$ are the antenna functions defined as follows \cite{finn1993observing}
\begin{align}\label{eq:antenna}
    &\mathrm{F_{+}=\frac{1}{2}(1+cos^2\theta)cos2\phi cos2\psi-cos\theta sin2\phi sin2\psi},\\\nonumber
     &\mathrm{F_{\times}=\frac{1}{2}(1+\cos^2\theta)\cos2\phi sin2\psi+cos\theta sin2\phi cos2\psi},
\end{align}
where $\rm{\theta}$ and $\rm{\phi}$ specify the source’s location in the detector-based coordinate system, and $\rm{\psi}$ is the polarization angle determining the orientation of the plus and cross modes with respect to the detector. (Note that while $\rm{i}$ and $\rm{\psi}$ both describe aspects of the binary’s orientation relative to the observer and the detector, they are independent parameters: $\rm{i}$  affects the intrinsic amplitude of the GW polarizations, whereas $\rm{\psi}$ affects the detector response.) Consequently, the matched filtering SNR ($\rho$) takes the form \cite{finn1996binary}

\begin{equation}
    \mathrm{\rho = \frac{\Theta}{4}\biggl[4\int_{f_{min}}^{f_{max}}h(f)^2/S_n(f)df\biggr]^{1/2}},
    \label{eq:SNR}
\end{equation}
where $\rm{\Theta^2 \equiv 4 \left(F_{+}^2(1+\cos^2i)^2 + 4F_{\times}^2\cos^2i\right)}$. Averaging over many binaries inclination angle and sky positions, \cite{finn1996binary} showed that $\Theta$ follows a distribution

\begin{equation}
\mathrm{P_{\Theta}(\Theta) = 
\begin{cases}
5\Theta(4-\Theta)^3/256& \text{if } 0<\Theta<4,\\
0,              & \text{otherwise}.
\end{cases}}
\end{equation}

In this study, we employ the next-generation GW detector combination of the CE and ET \cite{punturo2010einstein}. Specifically, we focus on the CE configuration with a 40 km arm length combined with ET, which we define as CEET. Although an SNR threshold of 8 is common for detection, we adopt \( \rm{\rho_{\rm Th} = 20} \) to select only golden events with high-precision parameter estimates. Events are simulated based on distances derived from redshifts, the parameter \( \rm{\Theta} \), and the mass distributions of the binary components. A redshift bin size of \( \rm{\Delta z = 0.05} \) is used, within which the total number of events is computed using Equation \eqref{eq:totno}. Assuming a local merger rate (\( \rm{R_0} \)) of \( \rm{20\,Gpc^{-3}\,yr^{-1}} \), and an operational period of five years with a 75\% duty cycle, the CEET system is projected to detect a significant number of events across a wide redshift range. Figure \ref{fig:EventCount} illustrates the projected total and detectable BBH, NSBH, and BNS events within specific redshift intervals (\( \rm{\Delta z = 0.05} \)). To obtain these estimates, the number of BBH, BNS, and NSBH mergers for each redshift bin is calculated using Equation \eqref{eq:totno}. For each merger, the masses of the binary components and the parameter \( \rm{\Theta} \) are determined using the inverse transform method, sampling from their respective probability distributions. Black hole masses are sampled within the range \( \rm{5M_{\odot}} \) to \( \rm{50M_{\odot}} \), while neutron star masses are sampled uniformly within \( \rm{1M_{\odot}} \) to \( \rm{2M_{\odot}} \). The parameter \( \rm{\Theta} \) is drawn from the interval \([0, 4]\). Using the redshift information for each event, the corresponding luminosity distance is calculated, and the SNR for a single detector is obtained using Equation \eqref{eq:SNR}. The total SNR, \( \rm{\rho_{\text{total}}} \), is then computed by combining the individual SNRs from the network of detectors using  $\rm{\rho_{\text{total}} = \sqrt{\sum_i \rho_i^2}}$. This analysis considers the detection capabilities of both the CEET network over a five-year observation period with a 75\% duty cycle. The curve depicted in figure \ref{fig:EventCount} shows the distribution of these events, providing insights into the temporal occurrence of these mergers across cosmic history and their detection likelihood with current and future GW observatories.  

\texttt{Parameter Estimation using Bilby:} We initiate the parameter estimation process for the identified sources using the \texttt{Bilby} package \cite{ashton2019bilby}, which provides realistic posterior distributions of the GW luminosity distance, marginalized over other source parameters. The masses of the GW sources and their number at different redshifts are determined according to methods described previously. For additional source parameters such as the inclination angle ($\rm{i}$), polarization angle ($\rm{\psi}$), and sky location (expressed in terms of right ascension (RA) and declination (Dec)), we adopt uniform sampling, assuming a non-spinning system (with RA and Dec obtained via standard coordinate transformations from the detector-based angles ($\rm{\theta}$) and ($\rm{\phi}$). We then generate a GW signal using the \texttt{IMRPhenomHM} waveform model \cite{kalaghatgi2020parameter}, which includes higher-order modes to help reduce the degeneracy between the luminosity distance ($\rm{D_L^{GW}}$) and the inclination angle ($\rm{i}$). By fixing the priors for all parameters except $\rm{m_1}$, $\rm{m_2}$, $\rm{D_L^{GW}}$, $\rm{i}$, RA, and Dec to delta-functions, we obtain detailed posterior distributions. Among these, the posterior distribution for $\rm{D_L^{GW}}$ is particularly vital for our analysis as it informs the inference of dark energy, while RA and Dec are essential for estimating sky localization errors.

\section{Redshift Inference of GW Sources}
\label{sec:RedInf}

An accurate distance and redshift measurements of GW sources are critical for inferring dark energy properties. While distance is determined robustly from the strain signal of GW sources, redshift inference varies depending on the nature of the source. For bright sirens, where an EM counterpart is identified, the redshift of the host galaxy can be directly and precisely measured through spectroscopic observations. These direct measurements are invaluable for constraining the dark energy EoS. However, for dark sirens, which lack an EM counterpart, redshift inference relies on statistical methods. 

Dark siren redshift inference is based on the spatial correlation between GW sources and galaxies. Since astrophysical black holes form in galaxies, which trace the large-scale distribution of dark matter, GW events are spatially correlated with galaxies. In the standard cosmological model, this correlation is characterized through bias parameters: $\rm{b_g(k, z)}$ for galaxies and $\rm{b_{GW}(k, z)}$ for GW sources. These parameters describe the relationship between the respective populations and the underlying dark matter density field. The three-dimensional auto-power spectra of each population and their cross-power spectrum are used to statistically infer the redshift distribution of GW sources. By cross-correlating the sky localization of GW events with galaxy surveys, a probabilistic redshift distribution for the host galaxies of GW sources can be constructed (see Section \ref{sec:CrossCor} for further details).

\subsection{Galaxy Surveys and GW Source Distribution}

In this analysis, we employ simulated catalogs from two major galaxy surveys: the Large Synoptic Survey Telescope (LSST) \citep{2009arXiv0912.0201L} and the Dark Energy Spectroscopic Instrument (DESI) \citep{DESI:2023ytc}. These surveys provide complementary strengths in redshift coverage and accuracy, making them particularly valuable for inferring redshift distributions of GW sources.

The LSST is a photometric survey with a catalog containing approximately three million galaxies extending up to redshifts of $\rm{z = 3}$. Its wide sky coverage, spanning nearly 18,000 square degrees ($\rm{f_{sky} \approx 0.436}$), and deep field observations enable the study of faint and distant objects. However, the reliance on photometric redshift estimates introduces larger uncertainties compared to spectroscopic measurements. In contrast, the DESI survey provides precise spectroscopic redshift measurements for approximately six million galaxies, with redshifts extending up to $\rm{z = 2}$. DESI covers about 14,000 square degrees of the sky ($\rm{f_{sky} \approx 0.339}$), offering highly accurate redshift determinations. However, its redshift range is more limited compared to LSST. Together, LSST and DESI play complementary roles in redshift inference. DESI's precise spectroscopic measurements are particularly advantageous for low-redshift GW sources detectable by current observatories. LSST's wider sky coverage and higher redshift reach make it well-suited for future GW detectors capable of observing more distant sources. By combining the strengths of these surveys, redshift inference can be enhanced across different epochs, enabling the inclusion of dark sirens in cosmological studies.

Both LSST and DESI catalogs are sourced from the CosmoHub database \cite{2017ehep.confE.488C,Tallada:2020qmg}. The galaxies in these catalogs are divided into tomographic bins with a redshift width of $\rm{\Delta z = 0.05}$. Gravitational wave sources are assigned to galaxies through random selection within these bins, assuming a homogeneous distribution of GW sources across the sky.

\subsection{Power Spectra Calculations and Bias Modeling}

For each redshift bin, we calculate the auto-power spectra of the galaxy and GW catalogs, as well as their cross-power spectrum, using the publicly available \texttt{Nbodykit} package~\cite{Hand:2017pqn}. These calculations are based on the non-linear matter power spectrum $\rm{P_m(k, z)}$, derived using Planck-2015 cosmology parameters \cite{Planck:2015fie}. The logarithmic growth function is set to zero for these calculations. The galaxy bias parameter $\rm{b_g(k, z)}$ is approximately constant at large scales ($\rm{k < 0.1 \, h/\text{Mpc}}$), with a value of $\rm{b_g \approx 1.6}$. Similarly, the GW source bias parameter $\rm{b_{GW}(k, z)}$ is expected to exhibit scale-independent behavior at large scales, but becomes scale-dependent at smaller scales due to processes such as binary formation, stellar metallicity, and feedback mechanisms. The redshift dependence of the GW bias can be modeled as $\rm{b_{GW}(z) = b_{GW}(1 + z)^\alpha}$, with $\rm{b_{GW} = 2}$ and $\rm{\alpha = 0}$ used in our analysis \cite{Mukherjee:2020mha, Diaz:2021pem}.

\section{Reconstruction of the Hubble Parameter}
\label{sec:HzReconstruction}

In this study, we investigate the measurement of the Hubble parameter, \(\mathcal{H}(z)\), using two classes of GW sources: bright sirens and dark sirens. Both source types provide critical insights into cosmology, with bright sirens offering direct redshift measurements via EM counterparts and dark sirens relying on statistical association techniques. Equation \ref{eq:DistRedshift} serves as the foundation for analyzing the expansion history of the Universe, correlating luminosity distance with redshift. The GW luminosity distance, \(D_L^{\text{GW}}\), is determined using \texttt{Bilby}, a Bayesian inference tool for GW parameter estimation.

To reconstruct \(\mathcal{H}(z)\) as a function of redshift, we employ a hierarchical Bayesian framework that incorporates uncertainties in both luminosity distance and redshift. The posterior probability density function of the Hubble parameter, given the data from \(n_{\text{GW}}\) GW sources, is expressed as
\begin{equation}
    \mathrm{P(\mathcal{H}(z)) \propto \Pi(\mathcal{H}(z)) \prod_{i=1}^{n_{\mathrm{GW}}} \iint  dD_L^{\mathrm{GW}^i} \, dz^i \, P(z^i) \mathcal{L}(D_L^{\mathrm{GW}^i} | \mathcal{H}(z^i), z^i),}
\end{equation}
where \(P(\mathcal{H}(z))\) represents the posterior probability density function of \(\mathcal{H}(z)\), \(\mathcal{L}(D_L^{\mathrm{GW}^i} \mid \mathcal{H}(z^i), z^i)\) is the likelihood of observing a given luminosity distance \(D_L^{\mathrm{GW}^i}\) given \(\mathcal{H}(z)\) and redshift \(z^i\), \(P(z^i)\) is the prior probability of the redshift, and \(\Pi(\mathcal{H}(z))\) is the prior on the Hubble parameter. A flat prior is chosen for \(\mathcal{H}(z)\) to ensure an unbiased exploration of the parameter space, enabling robust reconstruction without imposing strong constraints from prior knowledge. 

In the case of bright sirens, the dominant source of uncertainty arises from the luminosity distance error, which is significantly larger than the redshift error. Therefore, for bright sirens, \(P(z^i)\) is set as a delta function, corresponding to a precise measurement of the redshift. For dark sirens, \(P(z^i)\) is determined using Equation \ref{eq:LikeliCross}, which accounts for the statistical nature of redshift inference.

\begin{figure}[ht]
    \centering
    \includegraphics[height=8.0cm, width=16cm]{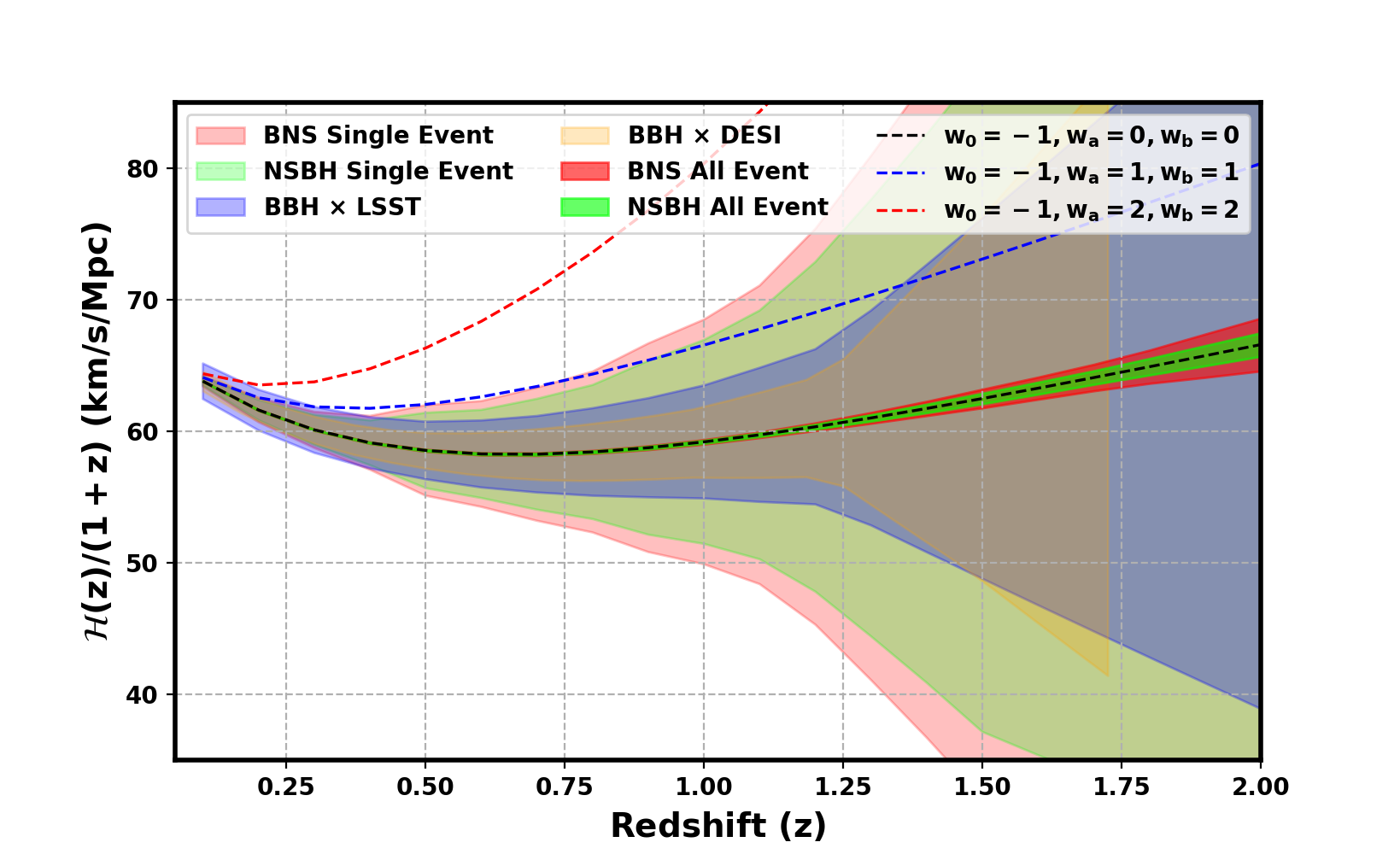}
    \caption{Reconstruction of the Hubble parameter \(\mathcal{H}(z)\) using gravitational wave sources for various scenarios. The plot includes results from single-event detections as well as the combined analysis of all sources for BNS and NSBH, where the redshift is directly inferred from electromagnetic counterparts. For BBH (dark siren), results are shown for two cases where redshift is inferred through cross-correlation with DESI and LSST galaxy surveys. The figure highlights the contributions of different source types and the role of galaxy surveys in improving the precision of \(\mathcal{H}(z)\) measurements. The dashed lines correspond to the fiducial values, which depend on the dark energy EoS as indicated in the legend.}
    \label{fig:HZRecon}
\end{figure}

In our analysis, we adopt a node-based reconstruction of \(\mathcal{H}(z)\). Specifically, we sample the Hubble parameter at a discrete set of redshift nodes, \(\{z_j\}\), with corresponding free parameters \(\mathcal{H}_j \equiv \mathcal{H}(z_j)\) (with \(j\) running over all the bins shown in Figure \ref{fig:EventCount}). After obtaining the posterior distributions for these node values, we interpolate between them to reconstruct the continuous function \(\mathcal{H}(z)\), which is then used to produce Figure \ref{fig:HZRecon}. This node-based method enables the data to directly inform the shape of \(\mathcal{H}(z)\) without assuming a specific functional form.

For the exploration of the free parameter space, we employ a Markov Chain Monte Carlo (MCMC) sampler implemented via the \texttt{emcee} Python package \citep{foreman2013emcee}. In our analysis, we assume that the likelihood for each GW source is Gaussian. For the \(i\)-th GW event, the likelihood function is written as
\begin{equation}
    \mathcal{L}\left(D_L^{\mathrm{GW},i} \mid \mathcal{H}(z^i), z^i\right) = \frac{1}{\sqrt{2\pi\,\sigma_{D_L^{GW},i}^2}} \exp\left[-\frac{\left(D_L^{\mathrm{GW},i} - D_L^{\mathrm{model}}(z^i; \mathcal{H}(z^i))\right)^2}{2\,\sigma_{D_L^{GW},i}^2}\right],
\label{eq:Likelihood}
\end{equation}
where \(D_L^{\mathrm{model}}(z^i; \mathcal{H}(z^i))\) is the theoretical luminosity distance computed from Equation \ref{eq:DistRedshift} and \(\sigma_{D_L^{GW},i}\) is the combined uncertainty on the luminosity distance measurement (including contributions from detector noise and weak lensing). The overall posterior distribution is then sampled using \texttt{emcee} to efficiently explore the parameter space defined by the node values \(\mathcal{H}_j\).

The error associated with \( \mathcal{H}(z) \) is primarily driven by two factors: the redshift error inferred from the cross-correlation of galaxy surveys and GW sources for dark sirens, and the luminosity distance error (\( D_L^{GW} \)) for both dark and bright sirens. The accuracy of luminosity distance measurements in GW astronomy is influenced not only by the sensitivity and configuration of the detector but also significantly by weak lensing, especially for sources detected at high redshifts by CEET. Gravitational lensing affects GW similarly to EM radiation. For GW events expected from large redshifts ($\rm{z > 1}$), weak lensing is common, along with occasional strongly-lensed sources. A lens with magnification $\rm{\mu}$ modifies the observed luminosity distance to $\rm{\frac{D_L^{GW}}{\sqrt{\mu}}}$, introducing a systematic error $\rm{\Delta D_L^{GW} / D_L^{GW} = 1 - \frac{1}{\sqrt{\mu}}}$ \cite{Mpetha:2022xqo}. This lensing error, when convolved with the expected magnification distribution $\rm{p(\mu)}$ from a standard $\Lambda$CDM model, significantly influences the overall measurement accuracy. In this study, we use the following model to estimate the error due to weak lensing \cite{Hirata:2010ba} 
\begin{equation}
    \mathrm{\frac{\sigma_{WL}}{D_L^{GW}} = \frac{0.096}{2} \left(\frac{1 - (1 + z)^{-0.62}}{0.62}\right)^{2.36}}.
\end{equation}
The total uncertainty in the measured luminosity distance is therefore expressed as
\begin{equation}
    \mathrm{\sigma_{D_L^{GW}}^2 = \sigma_{GW}^2 + \sigma_{WL}^2},
\end{equation}
where $\rm{\sigma_{GW}}$ is the uncertainty derived from the detector's setup during parameter estimation, and $\rm{\sigma_{WL}}$ accounts for the lensing effects. This comprehensive approach ensures a more accurate understanding of the intrinsic properties of GW sources. Another source of error in redshift inference is the contamination from peculiar velocities, which primarily affects GW sources at low redshifts (z $\leq$ 0.1), where the relative motion of galaxies due to local gravitational interactions introduces additional uncertainty in the inferred redshifts \cite{Mukherjee:2019qmm,nimonkar2024dependence}. However, this effect becomes negligible for higher-redshift sources (z $\geq$ 0.1) because the peculiar velocity contribution diminishes in comparison to the Hubble flow. Since the majority of GW events used in this study are detected at moderate to high redshifts, the impact of peculiar velocities on the overall error budget is minimal and does not significantly affect our conclusions.

Figure \ref{fig:EventCount} illustrates the number of detectable events as a function of redshift under the CEET configuration for all compact binary types. Using these simulated sources, we reconstruct \(\mathcal{H}(z)\) and present a comparative analysis in Figure \ref{fig:HZRecon}, showcasing results for all classes of compact binary sources: BNS, NSBH, and BBH. For bright sirens, such as BNS and NSBH systems with electromagnetic counterparts, the redshift is directly measured from the counterpart. In contrast, BBH events, which are unlikely to have associated electromagnetic counterparts, require redshift inference through cross-correlation, and for this analysis, we utilize LSST and DESI galaxy surveys. This approach demonstrates the complementary roles of bright and dark sirens in constraining \(\mathcal{H}(z)\) across different source types and redshift ranges.

\section{Reconstruction of the Dark Energy Equation of State}
\label{sec:DEosReconstructon}

In the previous section, we explored the measurement of the Hubble parameter, \(\mathcal{H}(z)\), using both bright sirens and dark sirens. In this section, we focus on reconstructing the dark energy EoS, \(w(z)\), using both model-independent and parametric approaches. Specifically, we demonstrate how \(w(z)\), as a function of redshift, can be inferred from observations of all types of GW sources. Additionally, we examine parametric reconstructions where the parameters of specific dark energy models are estimated as functions of redshift.

\begin{figure}[ht]
    \centering
    \includegraphics[height=8.0cm, width=16cm]{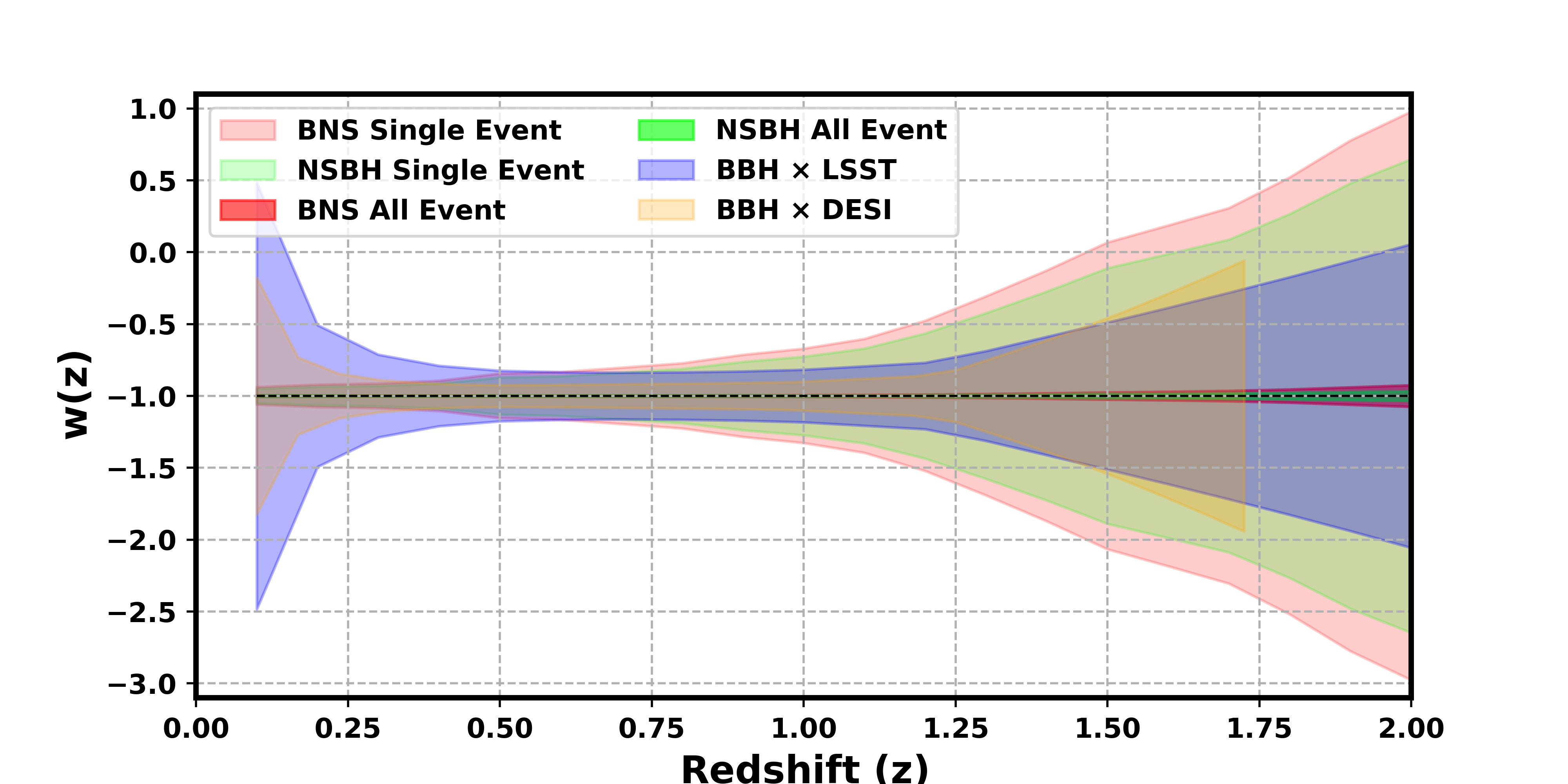}
    \caption{The figure presents the reconstructed \(w(z)\) for three types of compact binary sources: BNS, NSBH, and BBH. For BNS and NSBH systems with EM counterparts, redshifts are directly measured from the counterparts, with two cases analyzed: one with a single detected source and another with all sources having EM counterparts. For BBH events, redshifts are inferred through cross-correlation with galaxy surveys (LSST and DESI). This comparative analysis demonstrates the complementary roles of bright and dark sirens in constraining \(w(z)\) across different source types and redshift ranges.}
    \label{fig:DERecons}
\end{figure}

Firstly we  reconstruct the dark energy EoS parameter \(w(z)\) model independently using hierarchical Bayesian inference, we relate \(w(z)\) to the observed data, which consists of luminosity distances and redshifts from GW sources measurements of these observable are detailed in the Section \ref{sec:MockSamples} and \ref{sec:RedInf}. The posterior distribution for \(w(z)\) is expressed as:
\begin{equation}
    \mathrm{P(w(z)) \propto \Pi(w(z)) \prod_{i=1}^{n_{\mathrm{GW}}} \iint  dD_L^{\mathrm{GW}^i} \, dz^i \, P(z^i) \mathcal{L}(D_L^{\mathrm{GW}^i} | w(z^i), z^i),}
\end{equation}
where \(\mathrm{P(w(z))}\) is the posterior distribution of the equation of state parameter as a function of redshift, and \(\Pi(w(z))\) is the prior distribution on \(w(z)\), which can incorporate theoretical constraints or remain non-informative for a data-driven reconstruction. The product runs over the total number of GW sources, \(n_{\mathrm{GW}}\). The term \(P(z^i)\) represents the prior distribution of the redshift for the \(i\)-th GW source. For bright sirens, this prior is assumed to be a delta function, an approximation justified by the fact that the primary source of uncertainty arises from the luminosity distance measurement. In contrast, for dark sirens, the prior \(P(z^i)\) is determined using the cross-correlation technique, as detailed in Section \ref{sec:RedInf}.

The term \(\mathcal{L}(D_L^{\mathrm{GW}^i} | w(z^i), z^i)\) represents the likelihood of the observed luminosity distance \(D_L^{\mathrm{GW}^i}\) for the \(i\)-th source, conditioned on the dark energy equation of state \(w(z)\) and the source redshift \(z^i\). In our analysis, we assume that the likelihood for each GW source is Gaussian, as defined in Equation \ref{eq:Likelihood}. For the exploration of the free parameter space, we employ a MCMC sampler implemented via the \texttt{emcee} Python package \citep{foreman2013emcee}.

\begin{figure}[ht]
    \centering
    \includegraphics[height=11.0cm, width=12cm]{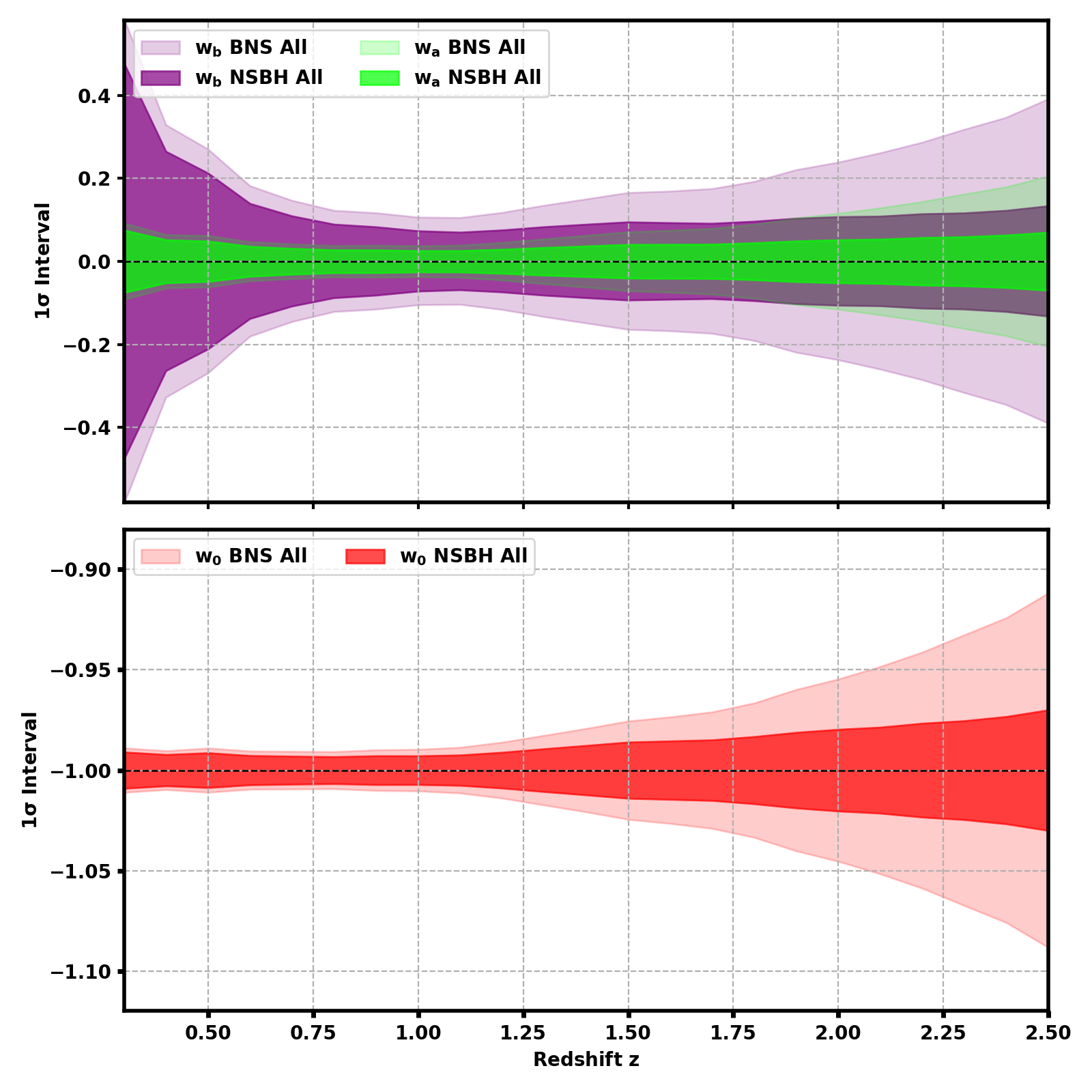}
    \caption{This plot showcases the precision achievable in reconstructing the dark energy equation of state parameters \(w_0\), \(w_a\), and \(w_b\) across redshift for  bright sirens, where the redshift is directly measured from electromagnetic counterparts.}
    \label{fig:DEParamRecons}
\end{figure}

\begin{figure}[ht]
    \centering
    \includegraphics[height=11.0cm, width=12cm]{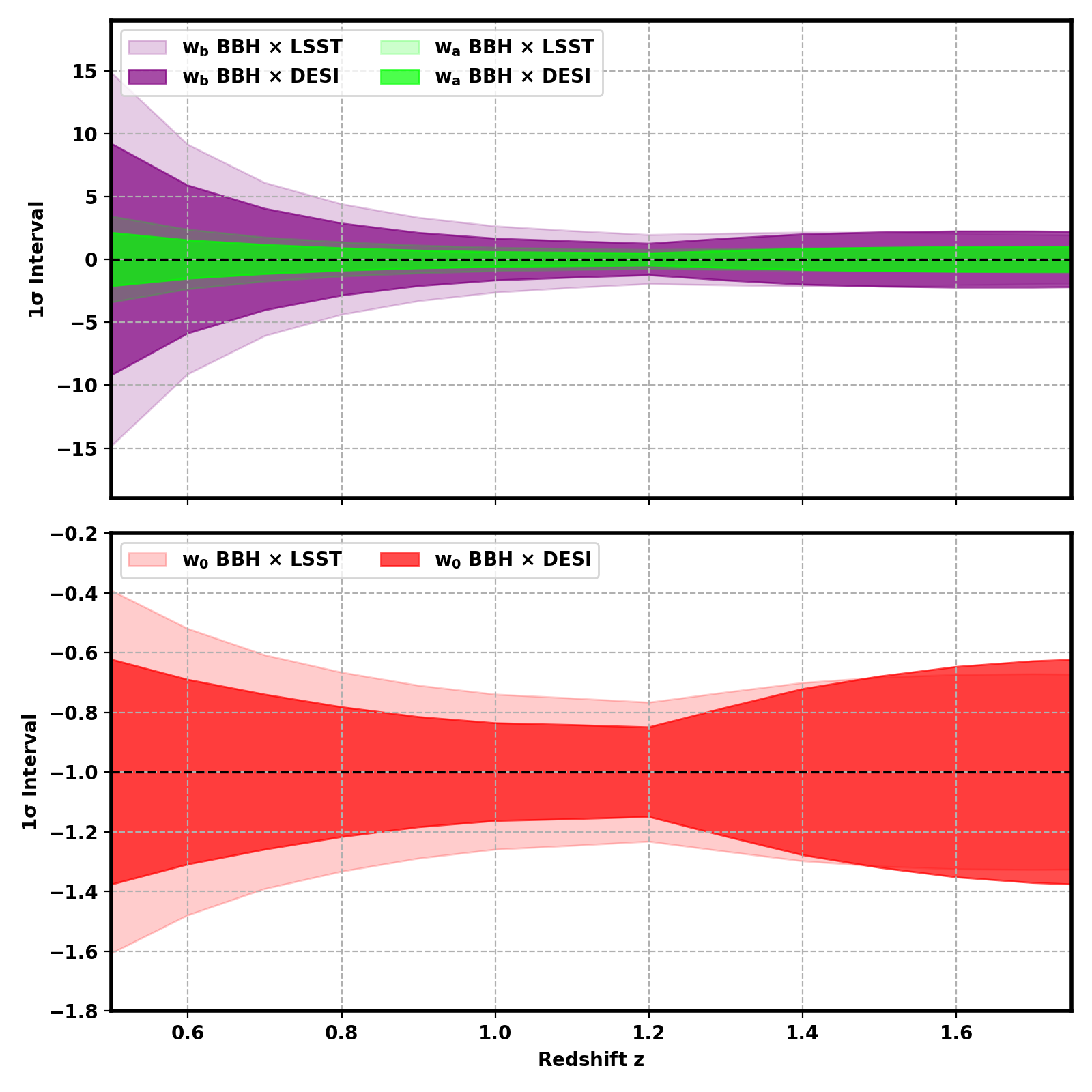}
    \caption{This plot showcases the precision achievable in reconstructing the dark energy equation of state parameters \(w_0\), \(w_a\), and \(w_b\) across redshift for dark sirens cross-correlated with DESI and LSST galaxy surveys.}
    \label{fig:DEParamReconsDarkSiren}
\end{figure}

Using the simulated sources mentioned in Section \ref{sec:MockSamples}, we reconstruct \(w(z)\) via a node-based approach, sampling \(w(z)\) at discrete redshift nodes and interpolating these values to obtain a continuous function, and present a comparative analysis in Figure \ref{fig:DERecons}, showcasing results for all classes of compact binary sources: BNS, NSBH, and BBH. For bright sirens, such as BNS and NSBH we analyze two cases: one where a single source is detected and another where all sources are detected. In contrast, BBH events, which are unlikely to have associated electromagnetic counterparts, require redshift inference through cross-correlation, and for this analysis, we utilize LSST and DESI galaxy surveys. This approach demonstrates the complementary roles of bright and dark sirens in constraining \(w(z)\) across different source types and redshift ranges.

To extend the parametric approach for reconstructing the dark energy EoS \(w(z)\), we generalize the commonly used CPL parametrization by introducing an additional term \(w_b\). This extended parametrization is defined in Equation \ref{eq:DEParametrization}, where \(w_0\) is the present-day value of the equation of state parameter, \(w_a\) describes its first-order redshift dependence, and \(w_b\) introduces a second-order correction to account for additional evolution at higher redshifts. This extension allows for greater flexibility in capturing the potential complexity of the dark energy behavior.

\begin{figure}[ht]
    \centering
    \includegraphics[height=13.0cm, width=14cm]{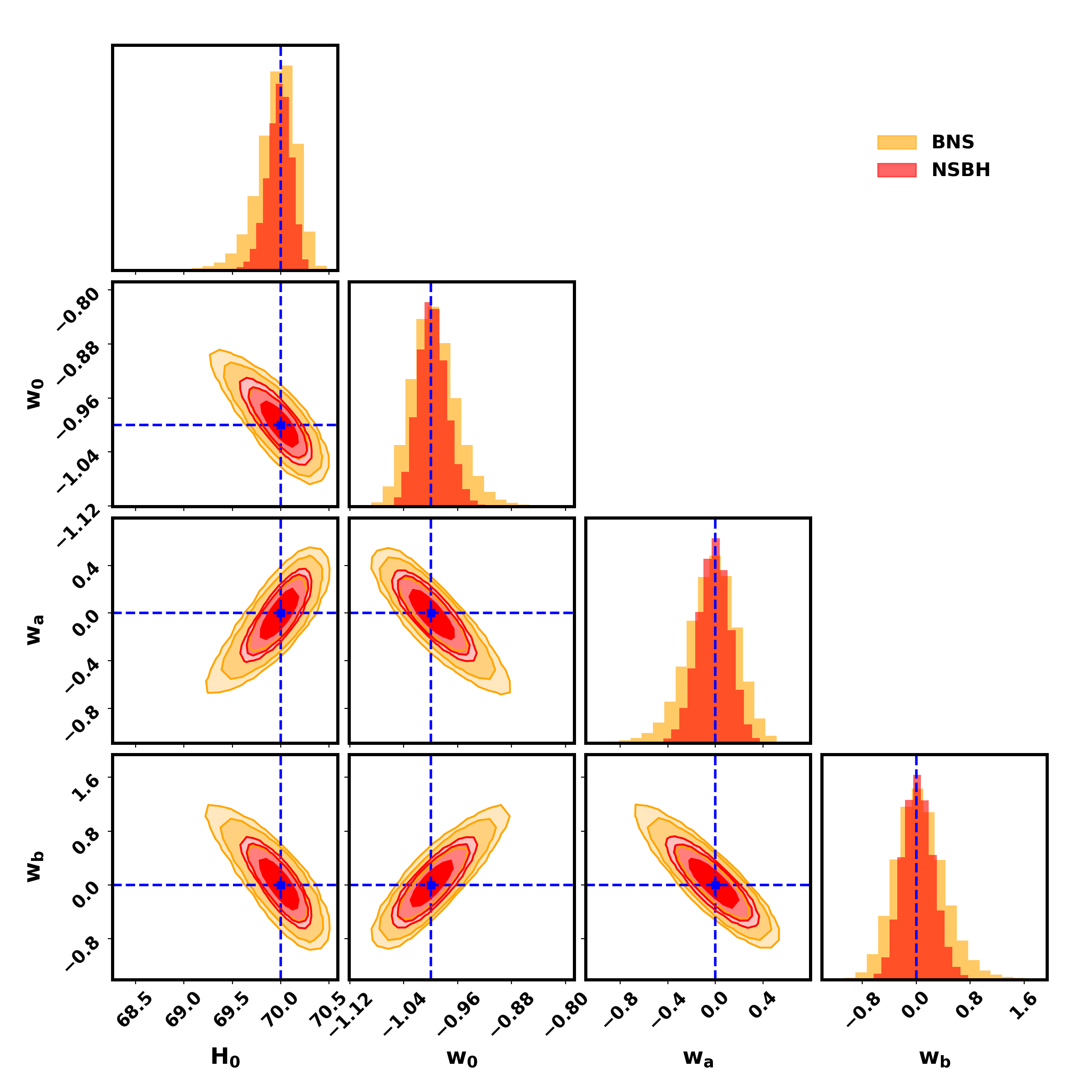}
    \caption{The figure presents the combined posterior distributions of the dark energy equation of state parameters $w_0$, $w_a$, and $w_b$, along with $H_0$, derived from all detected BNS and NSBH GW events. A total of 75,832 detected BNS and 152,054 detected NSBH events, distributed as shown in figure \ref{fig:EventCount}, are used in this analysis.}
    \label{fig:BrightsirensH0w0wawb}
\end{figure}

\begin{figure}[ht]
    \centering
    \includegraphics[height=13.0cm, width=14cm]{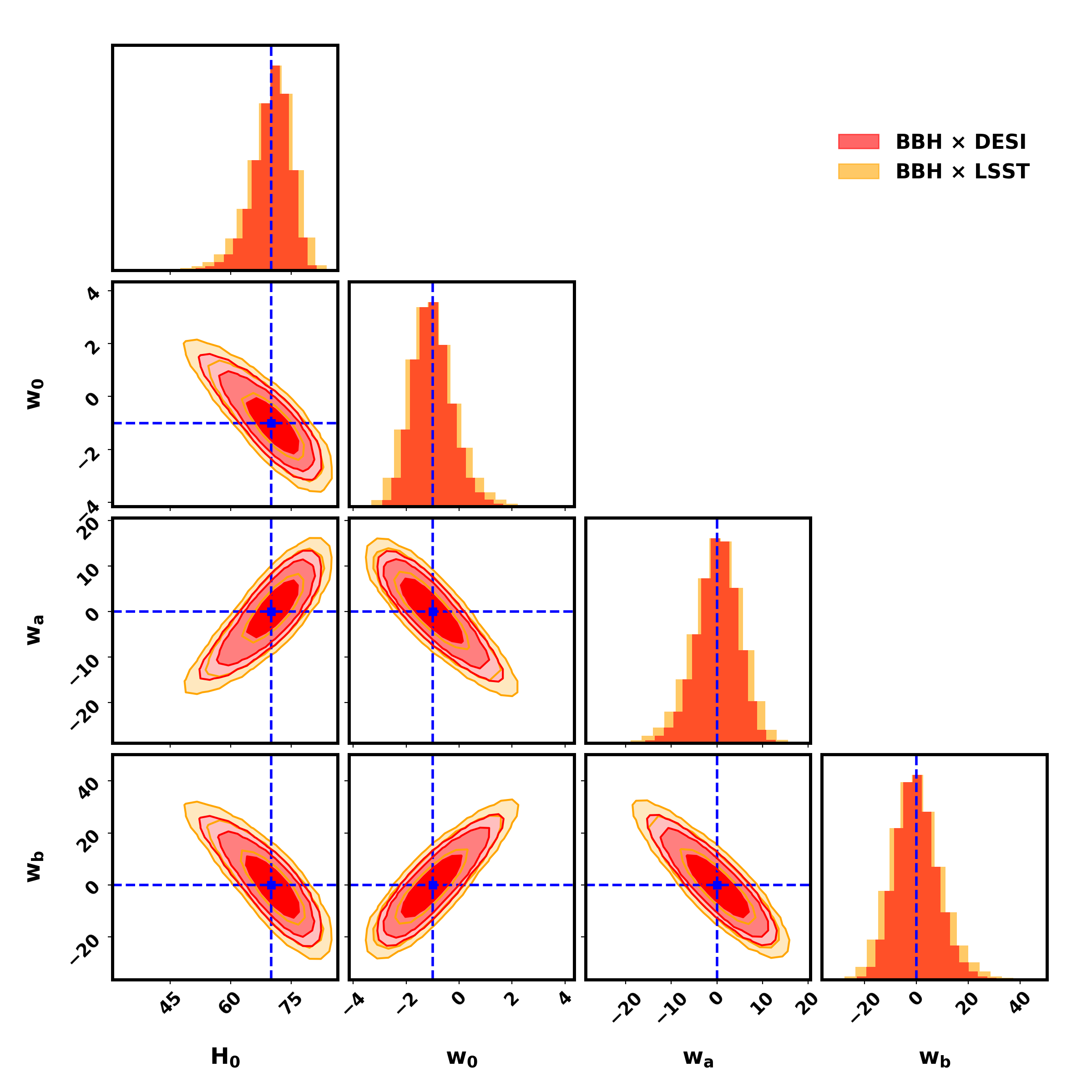}
    \caption{The figure presents the combined posterior distributions of the dark energy equation of state parameters, $w_0$, $w_a$, and $w_b$ along with $H_0$, derived from all detected BBH sources cross-correlated with the DESI and LSST galaxy surveys. A total of 180221 detected BBH events, distributed as shown in figure \ref{fig:EventCount}, are used in this analysis.}
    \label{fig:DarksirensH0w0wawb}
\end{figure}

\begin{figure}[ht]
    \centering
    \includegraphics[height=11.0cm, width=14cm]{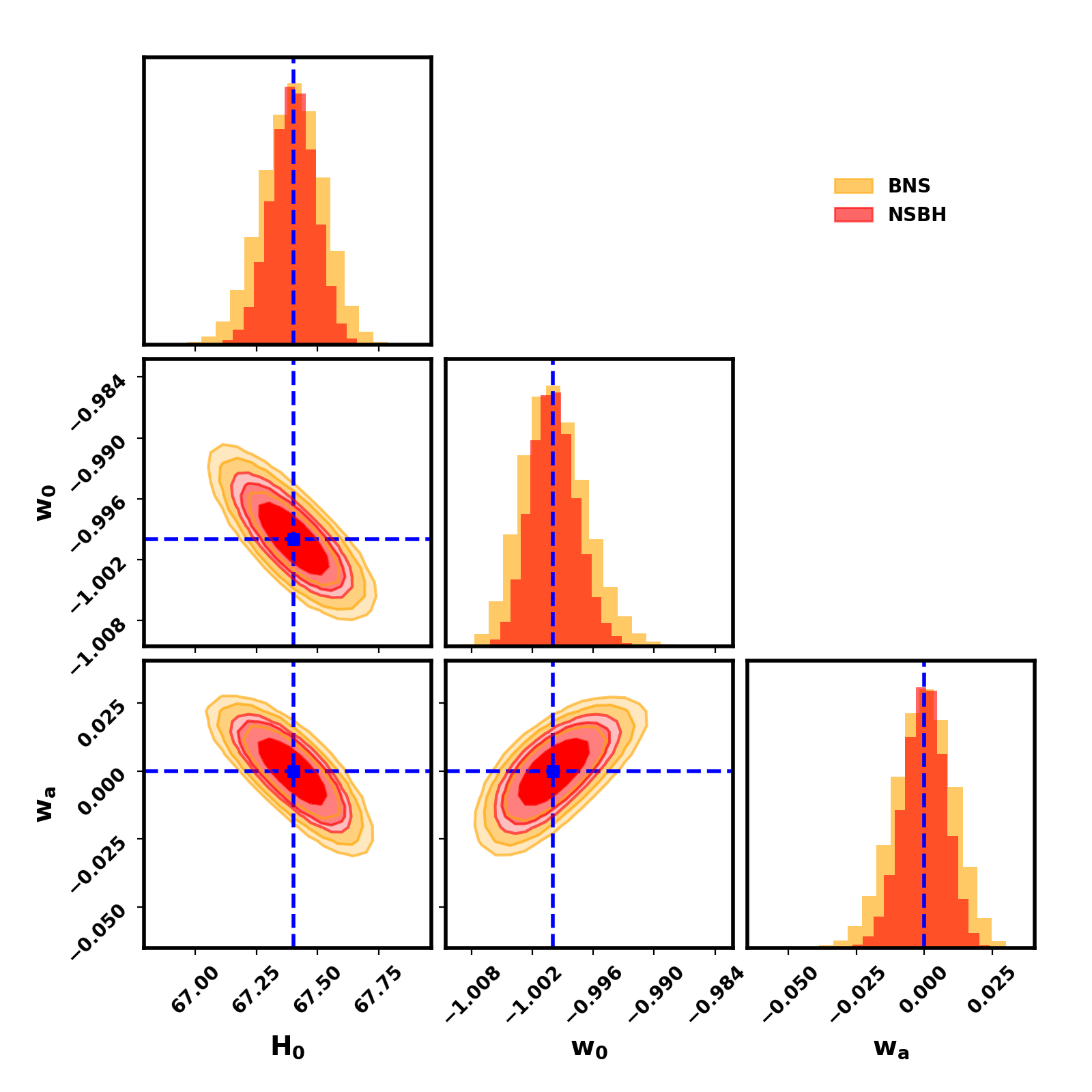}
    \caption{The figure shows the combined posterior distributions of the dark energy EoS parameters, $w_0$ and $w_a$, along with the Hubble constant, $H_0$, derived from all detected BNS and NSBH events. A total of 75,832 detected BNS and 152,054 detected NSBH events, distributed as shown in figure \ref{fig:EventCount}, are used in this analysis.}
    \label{fig:BrightsirensH0w0wa}
\end{figure}

\begin{figure}[ht]
    \centering
    \includegraphics[height=11.0cm, width=14cm]{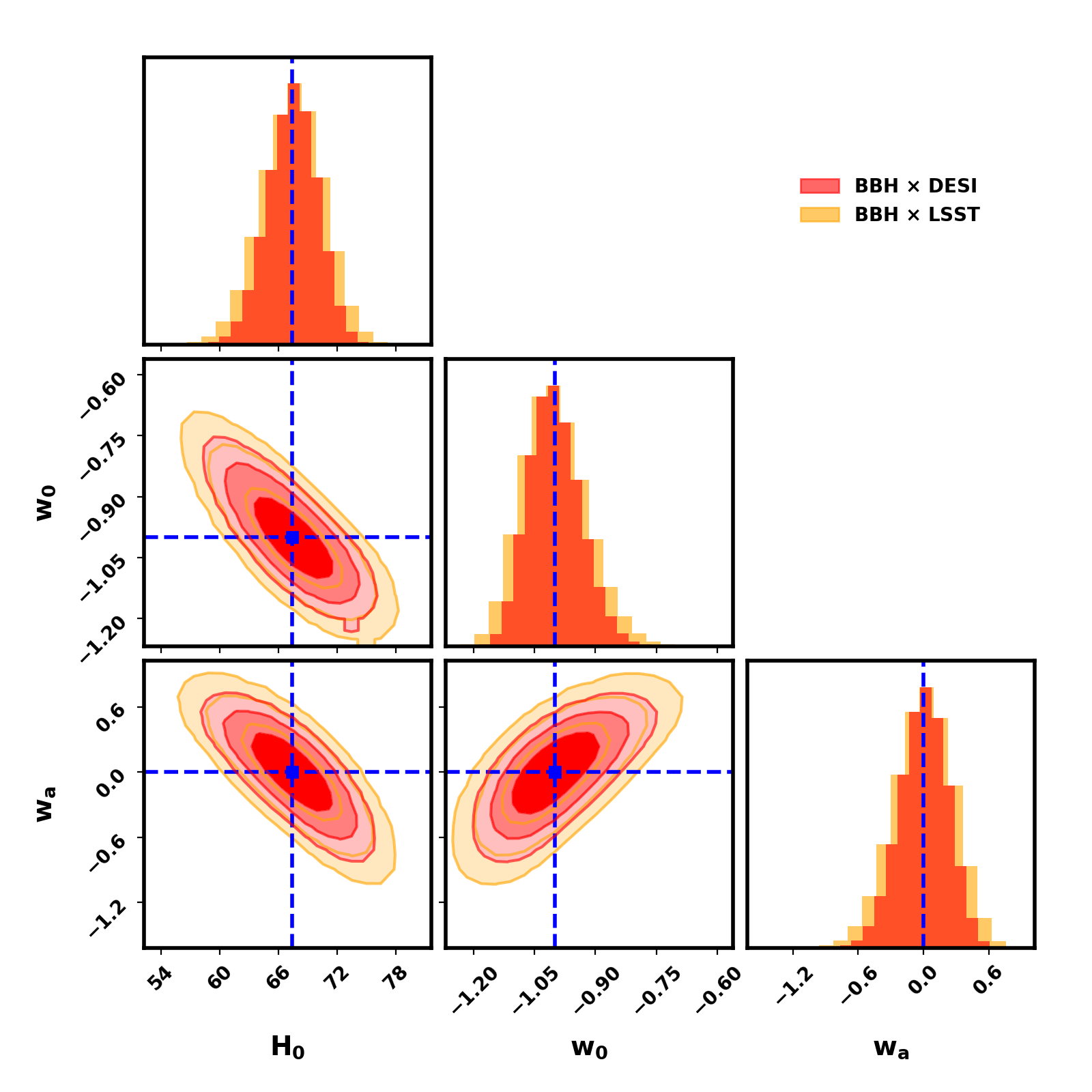}
    \caption{The figure shows the combined posterior distributions of the dark energy EoS parameters, $w_0$ and $w_a$, along with the Hubble constant, $H_0$, derived from all detected BBH sources cross-correlated with the DESI and LSST galaxy surveys. It highlights the constraints on these parameters, demonstrating the significant role of galaxy surveys in improving the measurement of $w(z)$ for dark sirens. A total of 180221 detected BBH events, distributed as shown in figure \ref{fig:EventCount}, are used in this analysis.}
    \label{fig:DarksirensH0wowa}
\end{figure}

Using this parametrization, we applied a hierarchical Bayesian framework to simultaneously infer the dark energy equation of state parameters \(w_0\), \(w_a\), and \(w_b\) along with Hubble constant $H_0$ from the observed luminosity distances and redshifts of GW sources. The posterior distribution for all four parameters is given by:

\begin{equation}
\begin{aligned}
    \mathrm{P}(H_0, w_0, w_a, w_b) \propto \Pi(H_0)\Pi(w_0)&\Pi(w_a)\Pi(w_b) \\ \nonumber &\times  \prod_{i=1}^{n_\mathrm{GW}} \iint dD_L^{\mathrm{GW}^i} \, dz^i \, P(z^i) \mathcal{L}\left(D_L^{\mathrm{GW}^i} \mid H_0, w_0, w_a, w_b, z^i\right),
\end{aligned}
\label{eq:Bayesian}
\end{equation}

here \(\mathrm{P}(H_0, w_0, w_a, w_b)\) represents the posterior distribution of the parameters, while \(\Pi(H_0)\), \(\Pi(w_0)\), \(\Pi(w_a)\), and \(\Pi(w_b)\) denote the prior distributions for the parameters \(H_0\), \(w_0\), \(w_a\), and \(w_b\), respectively. In this analysis, we have adopted flat priors for all parameters. The product in the expression runs over the total number of GW events, \(n_\mathrm{GW}\). The term \(P(z^i)\) refers to the prior on the redshift of the \(i\)-th GW source, while \(\mathcal{L}\) denotes the likelihood of observing the luminosity distance \(D_L^{\mathrm{GW}^i}\), given the parameter values \(H_0\), \(w_0\), \(w_a\), \(w_b\), and the redshift \(z^i\). In our analysis, we assume that the likelihood for each GW source is Gaussian, as defined in Equation \ref{eq:Likelihood}. For the exploration of the free parameter space, we employ a MCMC sampler implemented via the \texttt{emcee} Python package \citep{foreman2013emcee}. This parametric extension of the dark energy EoS enables us to probe deviations from the standard CPL form and test for more complex evolutionary behaviors in the context of GW cosmology.

\begin{figure}[ht]
    \centering
    \includegraphics[height=8.0cm, width=14cm]{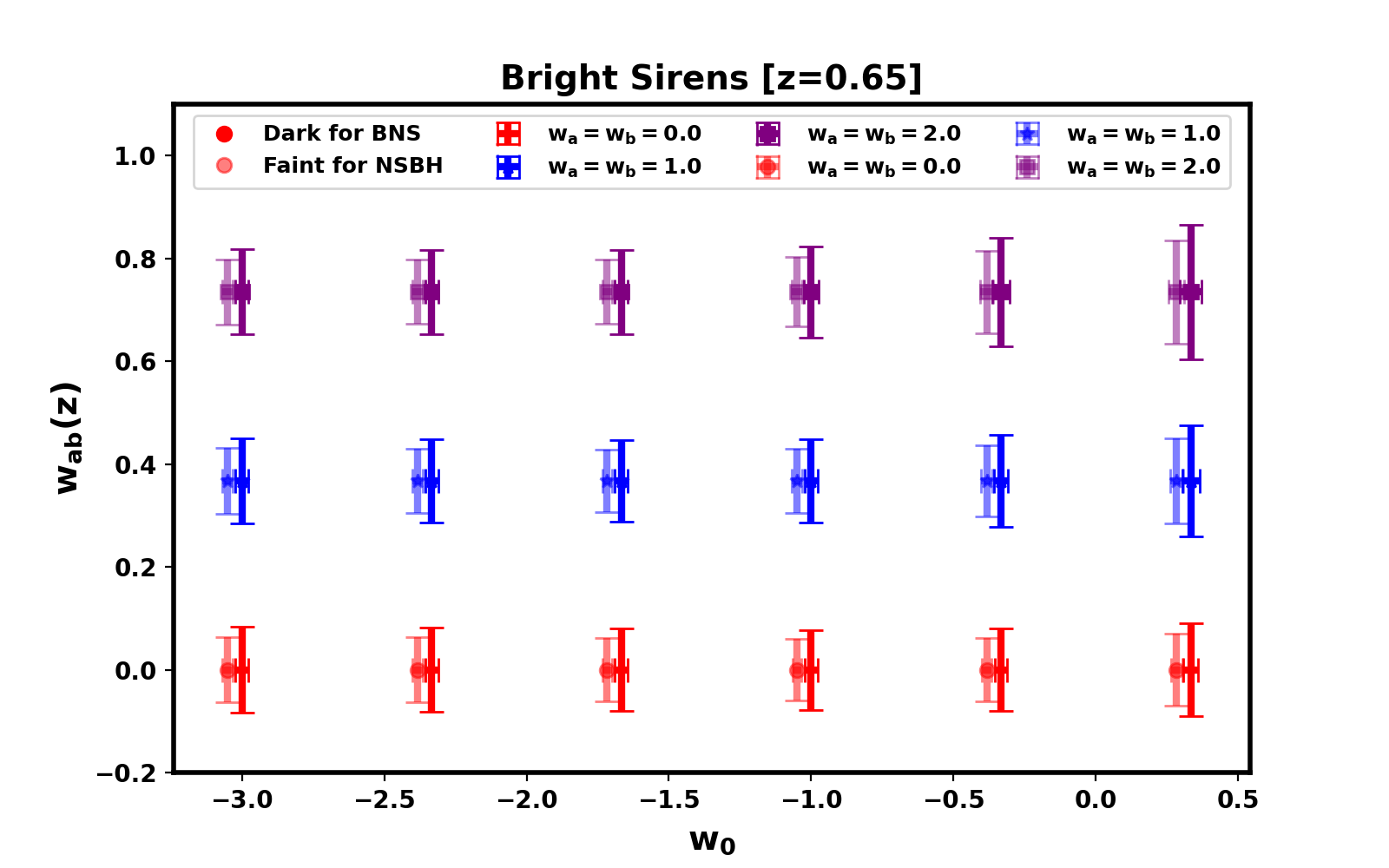}
    \caption{This plot illustrates the redshift-dependent component of the dark energy equation of state, \( w_{ab}(z) \), as a function of the fiducial value of \( w_0 \) at a redshift of \( z = 0.65 \), for the bright siren scenario. The fainter-colored curves represent the NSBH case, while the darker-colored curves correspond to the BNS case, highlighting how the uncertainty in \( w_{ab}(z) \) evolves with changes in \( w_0 \).}
    \label{fig:BrightSirenFinal}
\end{figure}

Using the simulated GW sources described in Section \ref{sec:MockSamples}, we reconstruct the dark energy EoS parameters \(w_0\), \(w_a\), and \(w_b\) in a redshift-dependent manner shown in Equation \eqref{eq:DEParametrization}. Specifically, for each redshift bin we use only the GW sources that fall within that bin to constrain the parameters. A comprehensive comparative analysis of these parameters is shown in Figures \ref{fig:DEParamRecons} and \ref{fig:DEParamReconsDarkSiren} for bright sirens and dark sirens, respectively. These figures depict the reconstructed values of \(w_0\), \(w_a\), and \(w_b\) across different redshift bins for all compact binary source classes. The analysis highlights the contributions of both bright and dark sirens in constraining the extended \(w_0\)-\(w_a\)-\(w_b\) parametrization of the dark energy EoS, demonstrating their sensitivity over a range of redshifts. Furthermore, the findings underscore the importance of GW observations in exploring the redshift evolution of dark energy.

For bright sirens, the uncertainties in the parameters $w_a$ and $w_b$ exhibit a characteristic behavior: they are relatively large at low redshifts, decrease to a minimum around $z \sim 1$, and then increase again at higher redshifts. This trend arises because of two effects: (i) at low redshifts, these parameters have minimal influence as the term $z/(1+z)$ is small. As redshift increases, the impact of $w_a$ and $w_b$ becomes more significant, leading to improved constraints. (ii) at redshifts around $z=1$, the number of sources are maximum. As a result, the error on the parameters $w_a$ and $w_b$ improves. However, at higher redshifts, the number of GW events diminishes, reducing the measurement accuracy for these parameters.

Similarly for $w_0$ the uncertainties are moderately high at low redshifts, decrease as the number of detected events increases, and then the uncertainty gradually increase again at higher redshifts due to lack of events. This behavior reflects the dependence of $w_0$ primarily on the overall number of events and their SNR. Additionally, NSBH binaries provide better constraints compared to BNS mergers. This is because the SNR of NSBH events is typically higher, owing to the greater mass of the black hole component. The redshift range where maximum number of sources will be detected from CE/ET will depend on the minimum delay time distribution. For this analysis we have considered a minimum delay time of 500 Myrs. If the minimum delay time increases or decreases, then the peak of the number of events will move to a lower or higher value, which will lead to a minimum uncertainty on the dark energy EoS parameters ($w_0,\, w_a,\, w_b$) at a lower or higher redshift than the fiducial case of minimum delay time of 500 Myrs considered in this analysis.

For dark sirens, the uncertainties in the parameters $w_a$ and $w_b$ are significantly larger at low redshifts due to the high uncertainty in redshift inference than the bright sirens. As redshift increases, the redshift determination improves, leading to a gradual reduction in uncertainties in $w_0, w_a$, and $w_b$, reaching a minimum around $z \sim 1.2$. Beyond this point, the uncertainties increase again, primarily due to the diminishing of GW events and the reduced precision of redshift measurements at higher redshifts from cross-correlation. The performance of redshift inference depends on the survey providing host galaxy information. DESI achieves better redshift accuracy up to $z \sim 1.5$ due to its spectroscopic redshift measurements, which are highly precise but limited in redshift reach. At higher redshifts, LSST becomes more effective, as its deep photometric surveys extend to fainter and more distant galaxies. This transition explains why the parameter uncertainties are influenced by the strengths and limitations of each survey across different redshift ranges.

In figure \ref{fig:BrightsirensH0w0wawb}, we present the joint inference of the dark energy EoS parameters \(w_0\), \(w_a\), and \(w_b\) along with $H_0$ for both BNS and NSBH systems. Similarly, figure \ref{fig:DarksirensH0w0wawb} shows the joint posterior distributions for dark sirens obtained through cross-correlation with the DESI and LSST galaxy surveys. These comparisons highlight the role of these surveys in improving redshift inference for dark sirens. Additionally, we present the joint inference for the standard CPL parametrization along with $H_0$ of dark energy EoS by setting \(w_b=0\) in figures \ref{fig:BrightsirensH0w0wa} and \ref{fig:DarksirensH0wowa}. This analysis includes both bright and dark sirens and incorporates all detectable GW events shown in figure \ref{fig:EventCount}. The values of the joint inference results for both the standard CPL and extended CPL parametrizations for fixed $H_0$ are summarized in Table \ref{tab:CombinedErrorSummary} and the correspoding plot are shown in Section \ref{sec:FixedH0Inf}. This table provides a detailed comparison of the inferred parameters, offering further insights into the sensitivity of GW observations to different dark energy models.

\begin{table}[h]
\renewcommand{\arraystretch}{1.4}
\centering
\begin{tabular}{|l|c|c|c|c|}
\hline

\multicolumn{5}{|c|}{\textbf{Fixed $H_0=67.4$ and $w_b=0$}} \\
\hline
Param & BNS & NSBH & BBH $\times$ DESI & BBH $\times$ LSST \\
\hline
$w_0$ & $-1.0^{+0.0027}_{-0.0027}$ & $-1.0^{+0.0016}_{-0.0017}$ & $-1.0^{+0.0500}_{-0.0480}$ & $-1.0^{+0.0630}_{-0.0630}$ \\
\hline
$w_a$ & $0.0^{+0.0099}_{-0.0099}$ & $0.0^{+0.0061}_{-0.0059}$ & $+0.003^{+0.1790}_{-0.1840}$ & $+0.001^{+0.2310}_{-0.2300}$ \\
\hline
\end{tabular}

\vspace{0.1cm}

\begin{tabular}{|l|c|c|c|c|}
\hline
\multicolumn{5}{|c|}{\textbf{Fixed $H_0=67.4$ with Varying $w_b$}} \\
\hline
Param & BNS & NSBH & BBH $\times$ DESI & BBH $\times$ LSST \\
\hline
$w_0$ & $-1.0^{+0.0100}_{-0.0100}$ & $-1.0^{+0.0060}_{-0.0060}$ & $-1.008^{+0.1870}_{-0.1880}$ & $-1.013^{+0.2340}_{-0.2380}$ \\
\hline
$w_a$ & $0.0^{+0.0800}_{-0.0800}$ & $0.0^{+0.0490}_{-0.0490}$ & $+0.069^{+1.4740}_{-1.4630}$ & $+0.111^{+1.8660}_{-1.8340}$ \\
\hline
$w_b$ & $0.0^{+0.1170}_{-0.1170}$ & $0.0^{+0.0720}_{-0.0720}$ & $-0.112^{+2.1510}_{-2.1810}$ & $-0.177^{+2.6880}_{-2.7660}$ \\
\hline
\end{tabular}
\caption{
Summary of the measured values of the dark energy parameters $w_0$, $w_a$, and $w_b$ under two scenarios. The top section represents results for a fixed $w_b = 0$, while the bottom section corresponds to a varying $w_b$, both derived using all detectable GW sources from three distinct categories: BNS, NSBH, and BBH mergers. For BBH sources, the results include cross-correlations with the DESI survey (BBH $\times$ DESI) and the LSST survey (BBH $\times$ LSST). 
The spatial distribution of these GW events is illustrated in figure \ref{fig:EventCount}. 
The table highlights the parameter constraints achieved for each source category, demonstrating the precision attainable with different GW source classes and redshift catalogs.}
\label{tab:CombinedErrorSummary}
\end{table}

\begin{figure}[ht]
    \centering
    \includegraphics[height=8.0cm, width=14cm]{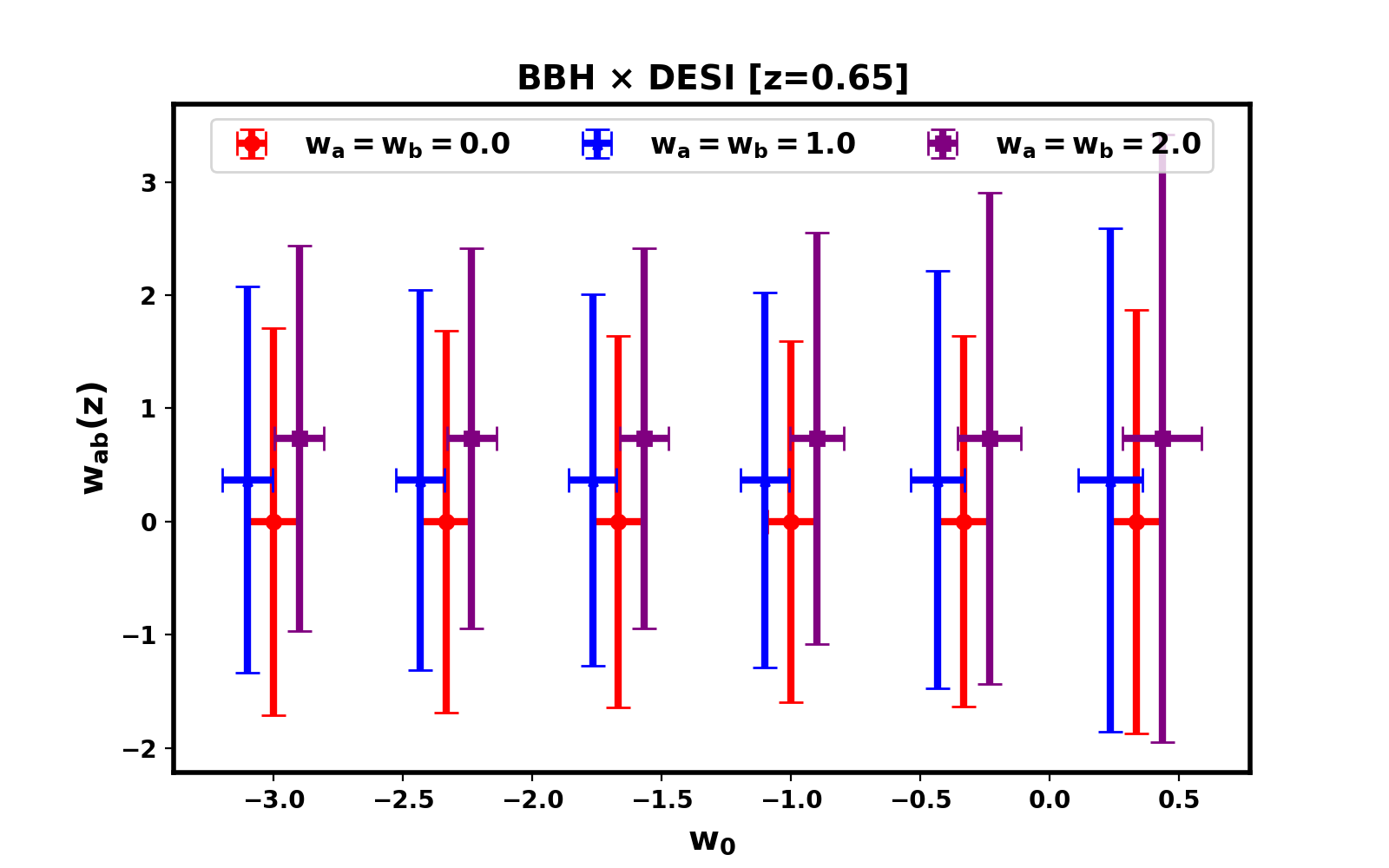}
    \caption{This plot illustrates the redshift-dependent component of the dark energy equation of state, \( w_{ab}(z) \), as a function of the fiducial value of \( w_0 \) at a redshift of \( z = 0.65 \). The analysis is based on the dark siren scenario cross-correlated with DESI, demonstrating how the uncertainty in \( w_{ab}(z) \) evolves with variations in \( w_0 \).}
    \label{fig:DarkSirenFinalDESI}
\end{figure}

\begin{figure}[ht]
    \centering
    \includegraphics[height=8.0cm, width=14cm]{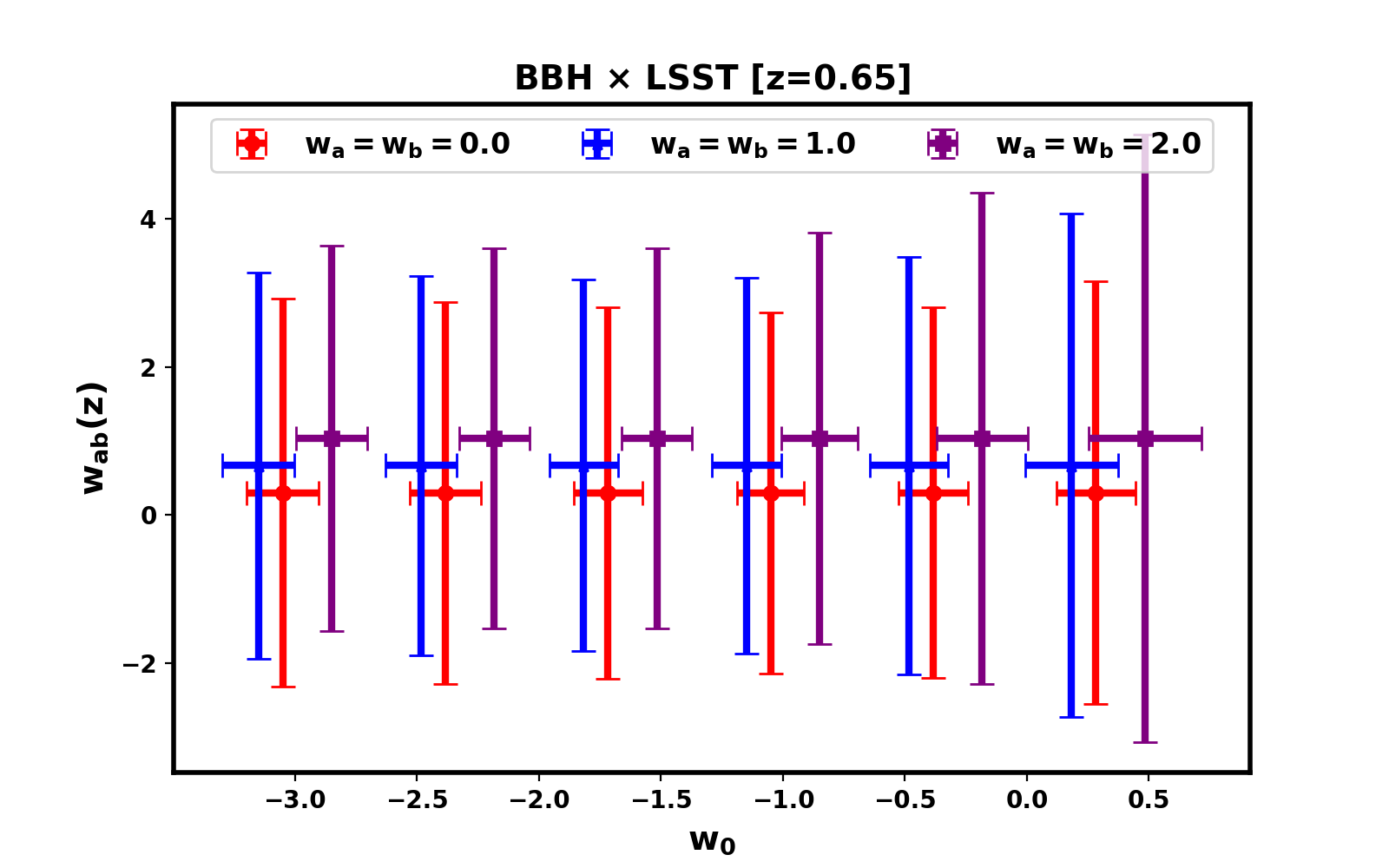}
    \caption{This plot illustrates the redshift-dependent component of the dark energy equation of state, \( w_{ab}(z) \), as a function of the fiducial value of \( w_0 \) at a redshift of \( z = 0.65 \). The analysis is based on the dark siren scenario cross-correlated with LSST, demonstrating how the uncertainty in \( w_{ab}(z) \) evolves with variations in \( w_0 \).}
    \label{fig:DarkSirenFinalLSST}
\end{figure}

Further to show the dependence of a fiducial choice of \( w_0 \) on the sensitivity in inferring the redshift-dependent part of the dark energy EoS $ w_{ab}(z)$ defined as 
\begin{equation}
    w_{ab}(z) = w_a \left(\frac{z}{1+z}\right) + w_b \left(\frac{z}{1+z}\right)^2,
\end{equation}
we estimate the uncertainty on the $w_0$ and $w_{ab}(z=0.65)$ for both bright and dark sirens. In figures \ref{fig:BrightSirenFinal}, \ref{fig:DarkSirenFinalDESI} and \ref{fig:DarkSirenFinalLSST}, we present how the measurement uncertainty of the redshift-dependent term, \( w_{ab}(z) \), varies as a function of the fiducial value of \( w_0 \) for bright and dark sirens, respectively at a redshift of $z=0.65$. These results demonstrate that the measurement precision of \( w_{ab}(z) \) is influenced by the choice of \( w_0 \). For bright sirens, we show results for both the BNS and NSBH cases. As expected, the NSBH measurements exhibit better accuracy compared to the BNS case, owing to the higher number of NSBH events and their higher SNR. For dark sirens, we show results for BBH events cross-correlated with DESI and LSST galaxy catalogs. As anticipated, at low redshifts, measurements using DESI outperform those using LSST due to DESI's superior redshift determination capabilities.

\begin{figure}[ht]
    \centering
    \includegraphics[height=8.0cm, width=14cm]{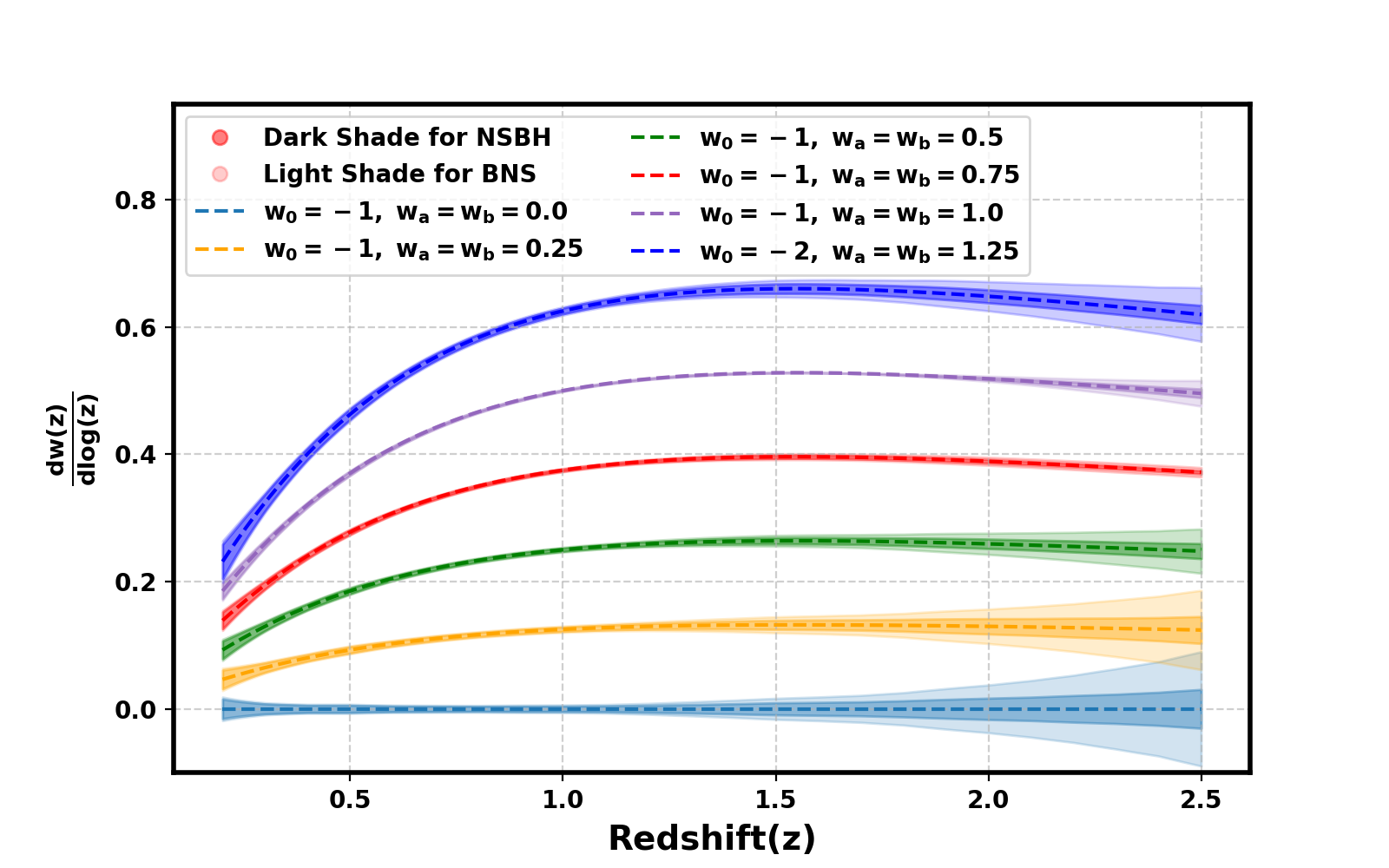}
    \caption{This plot illustrates the logarithmic derivative of the dark energy equation of state w(z) as a function of redshift z, under different parameterizations of the dark energy dynamics. The shaded regions correspond to uncertainties derived from two distinct GW source types: NSBH and BNS systems.}
    \label{fig:logplot}
\end{figure}

In order to understand the nature of dark energy it is crucial to measure the redshift evolution of the dark energy EoS which can test different theories of dark energy. In particular, the quantity \( \frac{dw(z)}{d\log(z)} \), which represents the rate of change of the EoS with respect to logarithmic redshift, provides a direct probe of how dark energy dynamics evolve over cosmic time. This measure is particularly valuable observationally, as it highlights subtle trends and transitions in the EoS that might otherwise be obscured in the analysis of \( w(z) \) alone. Theoretically, \( \frac{dw(z)}{d\log(z)} \) can distinguish between different dark energy models, including those that extend beyond the standard \( w_0 \)-\( w_a \) parametrization. In figure \ref{fig:logplot}, we present \( \frac{dw(z)}{d\log(z)} \) as a function of redshift for the BNS and NSBH cases, incorporating all the detected events. This quantity captures the rate of change of the dark energy EoS with respect to logarithmic redshift, offering valuable insights into the possible constraints that can be obtained from GW observations on the evolution of dark energy dynamics over cosmic time. The analysis is carried out for different values of the parameters \( w_0 \), \( w_a \), and \( w_b \), illustrating how variations in these parameters influence the temporal behavior of dark energy.

\begin{figure}[h]
    \centering
    \includegraphics[height=14.0cm, width=12cm]{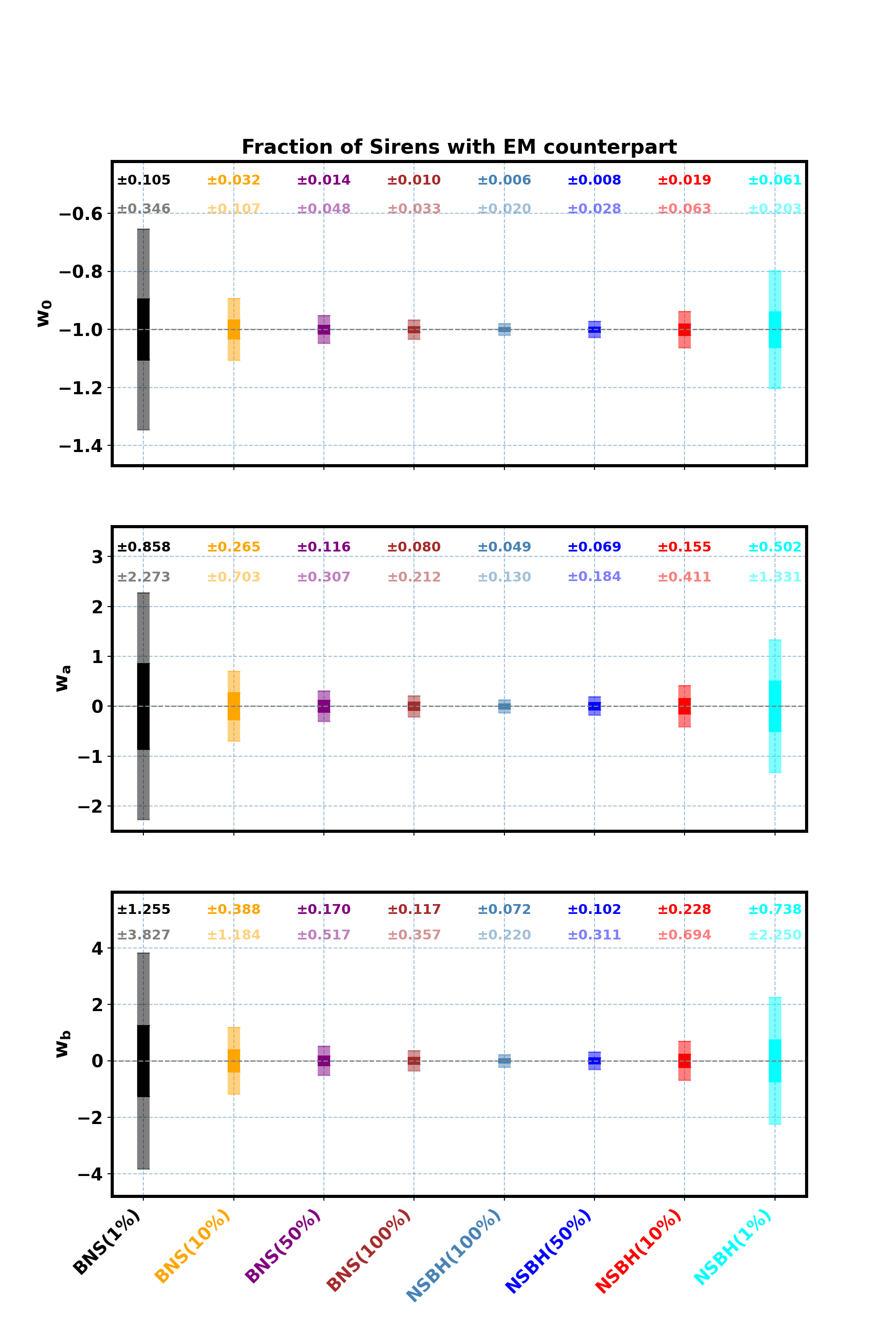}
    \caption{The figure illustrates the impact of varying fractions of GW events with EM counterparts on the measurements of the dark energy EoS parameters $w_0$, $w_a$, and $w_b$. The analysis considers scenarios with EM counterpart fractions of 1\%, 10\%, 50\%, and 100\%, for both a fixed value of Hubble constant $H_0 = 67.4$ km/s/Mpc (shown in darker shade) and a varying $H_0$ (shown in lighter shade). The constraints on these parameters improve significantly as the fraction of detected EM counterparts increases, demonstrating the importance of EM-GW multi-messenger observations for precision cosmology.}
    \label{fig:EMfractEffect}
\end{figure}

As the parameters \( w_a \) and \( w_b \) increase, the peak of \( \frac{dw(z)}{d\log(z)} \) shifts to higher redshifts and its magnitude increases, signifying a more pronounced evolution of the EoS at earlier epochs. The errors, represented by the shaded regions in the figure, are smallest at intermediate redshifts (\( z \sim 1.0 \)) and grow at both lower and higher redshifts. This pattern reflects the greater uncertainties in redshift determination due to limited detections and increased measurement errors for nearby and distant events. The trends observed in \( \frac{dw(z)}{d\log(z)} \) highlight its sensitivity to the choice of dark energy parameters. The distinct separation between the curves corresponding to different parameter sets emphasizes the potential of this diagnostic to constrain the dynamics of dark energy and differentiate between competing models. Furthermore, the minimal overlap in the shaded regions at intermediate redshifts suggests that this range offers the highest sensitivity for inferring the EoS parameters. By examining these variations and associated uncertainties, we can enhance our understanding of the evolution of dark energy across cosmic time.

In Figure \ref{fig:EMfractEffect}, we illustrate the impact of varying fractions of detectable EM counterparts on the measurement precision of the dark energy EoS parameters: $w_0$, $w_a$, and $w_b$. The analysis considers scenarios with EM counterpart fractions of 1\%, 10\%, 50\%, and 100\%, for both a fixed Hubble constant ($H_0 = 67.4  \, \mathrm{km \, s^{-1} \, Mpc^{-1}}$) and a varying $H_0$. The darker shades in the figure represent constraints for the fixed $H_0$ case, while lighter shades correspond to the varying $H_0$ scenario. As the fraction of detectable EM counterparts increases, the constraints on the EoS parameters become significantly tighter. This highlights the crucial role of multi-messenger observations in enhancing the precision of dark energy measurements. The results underscore the importance of combining GW detections with EM signals to advance our understanding of dark energy cosmology.

In Figure \ref{fig:SurveyComp} (Section \ref{sec:Introduction}), we present the key findings of our analysis along with a comparison to other upcoming cosmic probes of dark energy. In this figure, we compare the constraints derived from various combinations of GW observations including BNS, NSBH, and BBH systems, under the assumption of fixed values for $H_0 = 67.4 \, \mathrm{km \, s^{-1} \, Mpc^{-1}}$ and $w_b = 0$. For BBH events, redshift estimation is performed through cross-correlation with large-scale structure surveys, specifically DESI (BBH$\times$DESI) and LSST (BBH$\times$LSST). Additionally, we explored how the constraints from BNS and NSBH scenarios depend on the fraction of EM counterpart detections associated with these events. The figure illustrates four cases for  BNS and NSBH: the curve in lightest shade corresponds to only 1\% EM counterpart detection, followed by 10\%, 50\%, and the darkest curve representing 100\% EM counterpart detection. These results demonstrate that higher EM detection rates significantly enhance the precision, leading to tighter constraints on the dark energy EoS parameters. We further compared the performance of GW observations with other cosmological surveys, specifically WFIRST \citep{2013arXiv1305.5425S}, DESI \citep{DESI:2016fyo}, and Euclid \citep{Euclid:2021cfn}, using their forecasted constraints on the dark energy EoS parameters. For WFIRST and DESI, the error bars were visually extracted from \href{https://ui.adsabs.harvard.edu/abs/2021MNRAS.507.1746E/abstract}{figure 5} of \cite{Eifler:2020vvg} and \href{https://arxiv.org/abs/1611.00036}{figure 2.11} of \cite{DESI:2016fyo}, respectively, and are therefore treated as approximate values. The WFIRST constraints arise from a joint likelihood analysis that combines multiple cosmological probes, such as weak gravitational lensing, galaxy clustering, baryon acoustic oscillations, galaxy clusters, and Type Ia supernovae. Similarly, the DESI results incorporate measurements from the 14k BAO dataset, power spectrum constraints up to $k < 0.2$ \, h$/\text{Mpc}$, and complementary data from \textit{Planck} CMB observations and the Bright Galaxy Survey. For Euclid, the constraints were taken from \href{https://arxiv.org/abs/2105.09746}{Table 3} of \cite{Euclid:2021cfn}, corresponding to the $\Lambda$CDM fiducial model. These results combine simulated survey data with additional astrophysical measurements, such as variations in the fine-structure constant, along with local experimental constraints.

\section{Conclusion \& Discussion}
\label{sec:Summery}

In this study, we discuss the feasibility of measuring dark energy equation of state, \(w(z)\), across redshifts with an unprecedented precision and accuracy using both bright and dark standard sirens using the multi-messenger avenue. Our methodology incorporates both model-independent and parametric approaches, enabling us to explore the nature of dark energy over a broad redshift range. By leveraging GW sources, we utilize the complementary strengths of bright and dark sirens, integrating data from two major galaxy surveys, LSST and DESI, to reconstruct the evolution of \(w(z)\).

First, we investigated a model-independent approach, which reconstructs \(w(z)\) directly from observed luminosity distances and redshifts of GW sources without assuming a specific functional form for \(w(z)\). This flexible framework provides a powerful tool for testing the evolution of dark energy, as it avoids reliance on predefined assumptions about its behavior. The model-independent method allows us to extract insights directly from the data, offering an unbiased means of probing the properties of dark energy. In addition, we employed a parametric approach by extending the widely used CPL parametrization of the dark energy equation of state. While the standard CPL model, characterized by the parameters \(w_0\) and \(w_a\), has proven effective in describing a variety of cosmological behaviors, it is limited in capturing more intricate, redshift-dependent dynamics. To address this limitation, we introduced an additional parameter, \(w_b\), as defined in Equation \ref{eq:DEParametrization}. This extension enhances the model’s flexibility, allowing for the representation of more complex evolutionary behaviors of dark energy and providing a better fit to observational data, especially when deviations from the standard CPL model are evident. In addition to dark energy estimation, we also presented the reconstruction of the Hubble parameter across redshift.

Both model-independent and parametric approaches were applied to reconstruct \(w(z)\) using different types of compact binary sources: BNS, NSBH, and BBH. For bright sirens, such as BNS and NSBH systems with electromagnetic (EM) counterparts, redshift measurements from the counterparts provided precise constraints on the dark energy parameters. In contrast, BBH events, which are less likely to have EM counterparts, required redshift inference through cross-correlation with galaxy surveys. By incorporating data from LSST and DESI, we demonstrated the significant role these surveys play in enhancing the constraints on dark energy parameters and improving our understanding of its evolution. We showed that it will be possible to measure the dark energy parameters with remarkable precision: \( \sigma(w_0) = 0.010 \), \( \sigma(w_a) = 0.049 \), and \( \sigma(w_b) = 0.072 \) , under the assumption of fixed values for $H_0 = 67.4  \mathrm{km  s^{-1} Mpc^{-1}}$. This analysis assumes 5 years of observation time (\(T_{\mathrm{obs}}\)) with a duty cycle of 75\%, and the precision is expected to improve with observation time as \(T_{\mathrm{obs}}^{-1/2}\).

The error bars on dark energy EoS from multi-messenger technique can be an order of magnitude better than achievable from the many other cosmological surveys as shown in figure \ref{fig:SurveyComp}. The key advantage of this technique relies on the accurate inference of distance and redshift from bright sirens. As GW sources do not require to be standarized and one do not need a distance ladder, the error on measuring the cosmic evolution of dark energy is extremely robust.  Theoretical implications of these findings will be profound. They allow for rigorous testing of models such as quintessence, phantom energy, and modifications to general relativity, as well as novel dark energy scenarios  \citep{Shlivko:2024llw, PhysRevD.108.103519,2010PhRvD..82j3502H,2012PhRvD..85l3530S,2012JCAP...06..036S,Keeley:2019esp}. Through a precise reconstruction of $w(z)$ and $\mathcal{H}(z)$, our analysis breaks degeneracies between these models, offering a pathway to distinguish between different dark energy dynamics and their potential deviations from the standard $\Lambda$CDM model. This analysis, therefore, not only establishes a connection between GW observations and cosmological insights but also guides future theoretical and observational efforts to explore the nature of dark energy.

This approach holds significant promise for both bright and dark standard sirens in the mHz range, detectable by future space-based detectors such as LISA. To further improve the accuracy of these measurements, a high-quality spectroscopic survey is essential. Such a survey would provide precise spectroscopic redshifts for galaxies, minimizing the impact of photometric redshift errors and enhancing the overall accuracy of cosmological distance measurements. Furthermore, a full-sky survey and an increased number of galaxy sources are crucial to refining our measurements of the Hubble parameter $\mathcal{H}(z)$ and dark energy equation of state parameters. These advancements, coupled with improvements in GW source detection and analysis techniques, will significantly enhance our ability to study dark energy and its evolution. Moreover, the synergy between GW observations and EM counterparts embodies the power of multi-messenger astrophysics. Multi-messenger observations leveraging GW detections from facilities like LIGO, Virgo, Karga and future observatories such as the Einstein Telescope Cosmic Explorer and LISA, combined with spectroscopic and photometric observations from facilities like LSST, Euclid, DESI, WFIRST, and the Vera C. Rubin Observatory will play a pivotal role in enhancing our understanding of cosmological phenomena. These combined observations will provide precise redshift measurements, better constraints on GW sources, and a clearer picture of the dynamics of dark energy. To achieve this, advancements in GW detector networks, wide-field spectroscopic and photometric surveys, and improved analysis methodologies are essential. Together, these multi-messenger facilities will unlock new opportunities for studying dark energy, probing fundamental physics, and exploring the evolution of the universe.

\section*{Acknowledgements}
This work is part of the \texttt{⟨data|theory⟩ Universe-Lab}, supported by TIFR and the Department of Atomic Energy, Government of India. The authors express gratitude to the computer cluster of \texttt{⟨data|theory⟩ Universe-Lab} and the TIFR computer center HPC facility for computing resources. The authors are thankful to the Cosmic Explorer and Einstein Telescope collaboration for providing the noise specifications. This work has made use of CosmoHub, which has been developed by the Port d'Informació Científica (PIC), maintained through a collaboration of the Institut de Física d'Altes Energies (IFAE) and the Centro de Investigaciones Energéticas, Medioambientales y Tecnológicas (CIEMAT) and the Institute of Space Sciences (CSIC \& IEEC). CosmoHub was partially funded by the 'Plan Estatal de Investigación Científica y Técnica y de Innovación' program of the Spanish government, has been supported by the call for grants for Scientific and Technical Equipment 2021 of the State Program for Knowledge Generation and Scientific and Technological Strengthening of the R+D+i System, financed by MCIN/AEI/10.13039/501100011033 and the EU NextGeneration/PRTR (Hadoop Cluster for the comprehensive management of massive scientific data, reference EQC2021-007479-P) and by MICIIN with funding from European Union NextGenerationEU(PRTR-C17.I1) and by Generalitat de Catalunya. We acknowledge the use of the following packages in this work: Astropy \cite{robitaille2013astropy, price2018astropy}, Bilby \cite{ashton2019bilby}, Pandas \cite{mckinney2011pandas}, NumPy \cite{harris2020array}, Seaborn \cite{bisong2019matplotlib}, Scipy \cite{virtanen2020scipy}, Dynesty \cite{speagle2020dynesty}, emcee \cite{foreman2013emcee}, Nbodykit \cite{Hand:2017pqn} and Matplotlib \cite{Hunter:2007}.

\appendix
\section{Cross-Correlation Technique for Redshift Estimation of Dark Sirens}
\label{sec:CrossCor}

The distribution of matter on large scales in the Universe is generally considered to be statistically uniform and isotropic, consistent with the Copernican principle. This allows the large-scale galaxy distribution to be characterized using the galaxy density field, $\rm{\delta_g(r)}$, which is defined as

\begin{equation}
    \mathrm{\delta_g(r) = \frac{n_g(r)}{\bar{n}_g} - 1},
\end{equation}

where \(\rm{n_g(r)}\) represents the number density of galaxies at a given position \(\rm{r}\), while \(\rm{\bar{n}_g}\) denotes the average number density of galaxies. In the framework of the standard cosmological model, the spatial distribution of galaxies is understood to trace the underlying matter distribution in the Universe. This relationship can be expressed as a biased representation of the matter density field, \(\rm{\delta_m(k,z)}\), with the connection given by

\begin{equation}
    \mathrm{\delta_g(k,z) = b_g(k,z) \delta_m(k,z)},
\end{equation}

here \(\rm{b_g(k,z)}\) is the galaxy bias parameter, and \(\rm{\delta_g(k,z)}\) is the Fourier transform of the real-space galaxy density field \(\rm{\delta_g(r,z)}\). The galaxy bias parameter \(\rm{b_g(k,z)}\) describes how galaxies trace the distribution of dark matter. Galaxy catalogs from current and future surveys, including DES \cite{DES:2017myr}, DESI \cite{DESI:2016fyo}, Euclid \cite{2010arXiv1001.0061R}, LSST \cite{2009arXiv0912.0201L}, and SPHEREx \cite{SPHEREx:2018xfm}, will extend up to a redshift of \(z = 3\). These surveys, when combined, will cover nearly the entire sky, significantly increasing the overlap with GW sources. By providing extensive and detailed information on the galaxy distribution, these surveys will improve our ability to test cosmological models with higher precision and offer deeper insights into the large-scale structure of the Universe. Astrophysical GW events are expected to occur in galaxies and therefore follow their spatial distribution. This distribution is described by a bias parameter, \(\rm{b_{GW}(k, z)}\), which is distinct from the galaxy bias parameter, \(\rm{b_g(k, z)}\). The density field for GW sources in real space, \(\rm{\delta_{GW}(k, z)}\), is defined as

\begin{equation}
    \mathrm{\delta_{GW}(k, z) = b_{GW}(k, z) \delta_m(k, z)},
\end{equation}

here, \(\rm{b_{GW}(k, z)}\) describes how GW sources trace the large-scale structure of the Universe. However, GW sources are characterized by their luminosity distance (\(\rm{D_L^{GW}}\)) and sky localization \((\rm{\theta_{GW}, \phi_{GW})}\), introducing a sky localization error \(\rm{\Delta \Omega_{GW}}\). Sky localization error refers to the uncertainty in determining the precise position of a GW source in the sky. This error arises due to the inherent uncertainties in measuring the declination ($\rm{\theta_{GW}}$) and right ascension ($\rm{\phi_{GW}}$)  of the GW source, and it is defined as \cite{Singer:2015ema}

\begin{equation}
    \mathrm{\Delta\Omega_{GW} = \sin(\theta_{GW}) \sqrt{\sigma_{\theta_{GW}}^2 \sigma_{\phi_{GW}}^2 - \sigma_{\theta_{GW} \phi_{GW}}^2}}, 
\end{equation}
here $\rm{\sigma_{\theta_{GW}}}$ represents the error in $\rm{\theta_{GW}}$ (in radians), $\rm{\sigma_{\phi_{GW}}}$ is the error in $\rm{\phi_{GW}}$ (in radians), and $\rm{\sigma_{\theta_{GW} \phi_{GW}}}$ is the covariance between $\rm{\theta_{GW}}$ and $\rm{\phi_{GW}}$. The covariance term $\rm{\sigma_{\theta_{GW} \phi_{GW}}}$ is calculated from the joint distribution of samples of $\rm{\theta_{GW}}$ and $\rm{\phi_{GW}}$ using the formula
\begin{equation}
    \mathrm{\sigma_{\theta_{GW} \phi_{GW}} = \frac{1}{N - 1} \sum_{i=1}^{N} (\theta_{GW_i} - \bar{\theta}_{GW})(\phi_{GW_i} - \bar{\phi}_{GW})}, 
\end{equation}
here $\rm{N}$ is the number of samples, $\rm{\theta_{GW_i}}$ and $\rm{\phi_{GW_i}}$ are individual samples of $\rm{\theta_{GW}}$ and $\rm{\phi_{GW}}$, and $\rm{\bar{\theta}_{GW}}$ and $\rm{\bar{\phi}_{GW}}$ are the means of the sample sets for $\rm{\theta_{GW}}$ and $\rm{\phi_{GW}}$, respectively. This error leads to uncertainty in the precise position of the GW source within the sky localization region, blurring the spatial information. Consequently, the density field of GW sources is modified due to this sky localization error, expressed as
\begin{equation}
    \mathrm{\delta_{\text{GW}}^r(k, \Delta \Omega_{GW}, z) = \delta_{\text{GW}}^r(k, z) e^{-\frac{k^2}{k_{\text{eff}}^2(z)}}},
\end{equation}
where \(\rm{\delta^r_{\text{GW}}(k, z)}\) is the original density field of GW sources without sky localization impact, and \(\rm{k_{\text{eff}}(z)}\) denotes a characteristic wavenumber varying with redshift \(\rm{z}\) \cite{Mukherjee:2020hyn,Mukherjee:2020mha}. \(\rm{k_{\text{eff}}(z)}\) is the comoving scale defined as 
\begin{equation}
    \mathrm{k_{\text{eff}}(z) = \sqrt{\frac{8 \ln 2}{\Delta \Omega_{GW} D_c(z)^2}}},
\end{equation}
where \(\rm{D_c(z)}\) is the comoving distance at redshift $\rm{z}$ \cite{Mukherjee:2020hyn}. The exponential term \(\rm{e^{-\frac{k^2}{k_{\text{eff}}^2(z)}}}\) quantifies the smoothing effect caused by the sky localization error, with a rapid decrease beyond \(\rm{k > k_{\text{eff}}}\).

\subsection{Mathematical Formulation of Cross-Correlation Technique}

To estimate the redshift ($\rm{z}$) of a GW binary source when its EM counterpart is not detectable, cross-correlation between galaxy surveys and dark sirens is utilized. The probability distribution for this estimation is expressed as:

\begin{equation}
   \mathrm{\mathcal{P}(z|\vec{\vartheta}_{GW}, \vec d_{g})\propto \mathcal{L}(\vec{\vartheta}_{GW}| P^{ss}_{gg}(\vec k,z), \vec d_g(z)) \mathcal{P}(\vec d_g| P^{ss}_{gg}(\vec k,z))},
\end{equation}

In this expression, $\rm{\vec{\vartheta}_{\text{GW}} = (D_L^{\text{GW}}, \theta_{\text{GW}}, \phi_{\text{GW}})}$ represents the GW data vector, which includes the luminosity distance $\rm{D_L^{GW}}$ to the GW source, as well as the sky localization of the source $(\theta_{\text{GW}}, \phi_{\text{GW}})$. The galaxy data vector $\rm{\vec{d}_g = (\delta_g(z_g, \theta_g, \phi_g))}$ includes the redshift information of the galaxy $\rm{z_g}$ as well as the sky position $(\rm{\theta_g, \phi_g})$. The term  $\rm{\mathcal{L}(\vec{\vartheta}_{\text{GW}} | P^{ss}_{gg}(\vec{k}, z), \vec{d}_g(z))}$ represents the likelihood function, which represents how well the GW data fits the galaxy density field based on a model for the galaxy power spectrum  $\rm{P^{ss}_{gg}(\vec{k}, z)}$. This power spectrum captures the distribution of galaxy densities in redshift space, accounting for both cosmological redshift and additional redshift effects due to the peculiar velocities of galaxies. The term $\rm{\mathcal{P}(\vec{d}_g | P{gg}^{ss}(\vec{k}, z))}$ denotes the posterior distribution of the galaxy density field given the power spectrum $\rm{P^{ss}_{gg}(\vec{k}, z)}$. It provides the probability of observing the galaxy data under the assumed model for the galaxy power spectrum. The superscript 's' indicates that the galaxy survey in question is either a spectroscopic or photometric survey, observing galaxies in redshift space, which includes both cosmological redshift and redshift due to peculiar velocities. Overall, the probability distribution,  $\rm{\mathcal{P}(z | \vec{\vartheta}_{GW}, \vec{d}_g)}$  combines information from GW data and galaxy surveys to estimate the redshift of the GW source. The posterior distribution of the galaxy density field, conditional on the galaxy power spectrum, is described by the equation

\begin{equation}
\mathrm{\mathcal{P}(\vec d_g| P^{ss}_{gg}(\vec k,z)) \propto \exp\bigg(-{\frac{ \delta^{s}_g(\vec k, z) \delta^{s*}_g(\vec k, z)}{2(P^{ss}_{gg}(\vec k,z) + n_g(z)^{-1})}}\bigg)},
\end{equation}
where $\rm{P^{ss}_{gg}(\vec{k}, z) = b^2_g(k, z) (1 + \beta_g \mu^2{\hat{k}})^2 P_m(k, z)}$ is the three-dimensional power spectrum of galaxies, $\rm{b_g(k, z)}$ is the galaxy bias parameter, $\rm{\beta_g = \frac{f}{b_g}}$ with $\rm{f \equiv \frac{d \ln D}{d \ln a}}$ describing the logarithmic growth function, $\rm{\mu_k = \cos(\hat{n} \cdot \hat{k})}$ is the cosine of the angle between the line of sight $\rm{\hat{n}}$ and the Fourier modes $\rm{\hat{k}}$, and $\rm{n_g(z) = \frac{N_g(z)}{V_s}}$ is the number density of galaxies within the redshift bin $\rm{z}$. The Fourier transform of the galaxy distribution, $\rm{\delta^s_g(\vec{k}, z)}$, is computed as $\rm{\int d^3\vec{r} , \delta_g(\vec{r}) e^{i\vec{k} \cdot \vec{r}}}$. Following the description of the probability distribution and posterior distribution, the likelihood function, $\rm{\mathcal{L}(\vec{\vartheta}_{\text{GW}} | P{gg}(\vec k,z), \Theta_n, \vec d_g(z))}$, is defined in Equation\eqref{eq:LikeliCross}
\begin{equation}
  \mathrm{\mathcal{L} \propto \exp\bigg(-\frac{V_s}{4\pi^2}\int k^2 dk \int d\mu_k
  \frac{ \bigg(\hat{P} (\vec{k}, \Delta \Omega_{GW}) - b_g(k,z)b_{GW}(k, z)(1 + \beta_g \mu_{\hat{k}}^2)P_{m}(k,z)e^{-\frac{k^2}{k^2_{\rm eff}}}\bigg)^2}
  {2\big(P^{ss}_{gg}(\vec{k},z) + n_g(z)^{-1}\big)\big(P^{rr}_{GW\,GW}(\vec{k},z) + n_{GW}(z)^{-1}\big)}
  \bigg),}
\label{eq:LikeliCross}
\end{equation} 
where $\rm{\hat{P}(\vec k, z)= \delta_{g}(\vec k, z)\delta_{\text{GW}}^*(\vec k,\Delta \Omega_{\text{GW}})}$ represents the cross power spectrum between the galaxy and GW data. The term $\rm{n_{\text{GW}}(z)= \frac{N_{\text{GW}}(D^i_l(z))}{V_s}}$ is the number density of gravitational wave sources described in terms of the number of objects within a specific luminosity distance bin. $\rm{V_s}$ denotes the total sky volume, and $\rm{P^{rr}_{\text{GW,GW}}(\vec k, z) = b^2_{\text{GW}}(k, z)P_m(k, z)}$ with $\rm{b_{\text{GW}}(k, z)}$ being the GW bias parameter. This formulation captures the cross-power spectrum analysis necessary for determining the correlation between galaxy distributions and GW sources, considering biases and noise properties, with the exponential term quantifying the fit discrepancy between observed and theoretical power spectra.

\section{Constraints on Dark Energy Equation of State with Fixed Hubble Constant}
\label{sec:FixedH0Inf}

In this section, we present the constraints on the dark energy EoS parameters under the assumption of a fixed Hubble constant value of \( H_0 = 67.4 \, \mathrm{km \, s^{-1} \, Mpc^{-1}} \) \cite{Planck:2018vyg}. The hierarchical Bayesian framework defined in Equation \ref{eq:Bayesian} is used for this analysis, and the inferred parameter values are summarized in Table \ref{tab:CombinedErrorSummary}. The aim of this section is to visualize the impact of fixing \( H_0 \) on the reconstructed dark energy EoS. The methods and frameworks used in the analysis have been thoroughly discussed in earlier sections, and no new theoretical developments are introduced here.

\begin{figure}[ht]
    \centering
    \includegraphics[height=7.65cm, width=10cm]{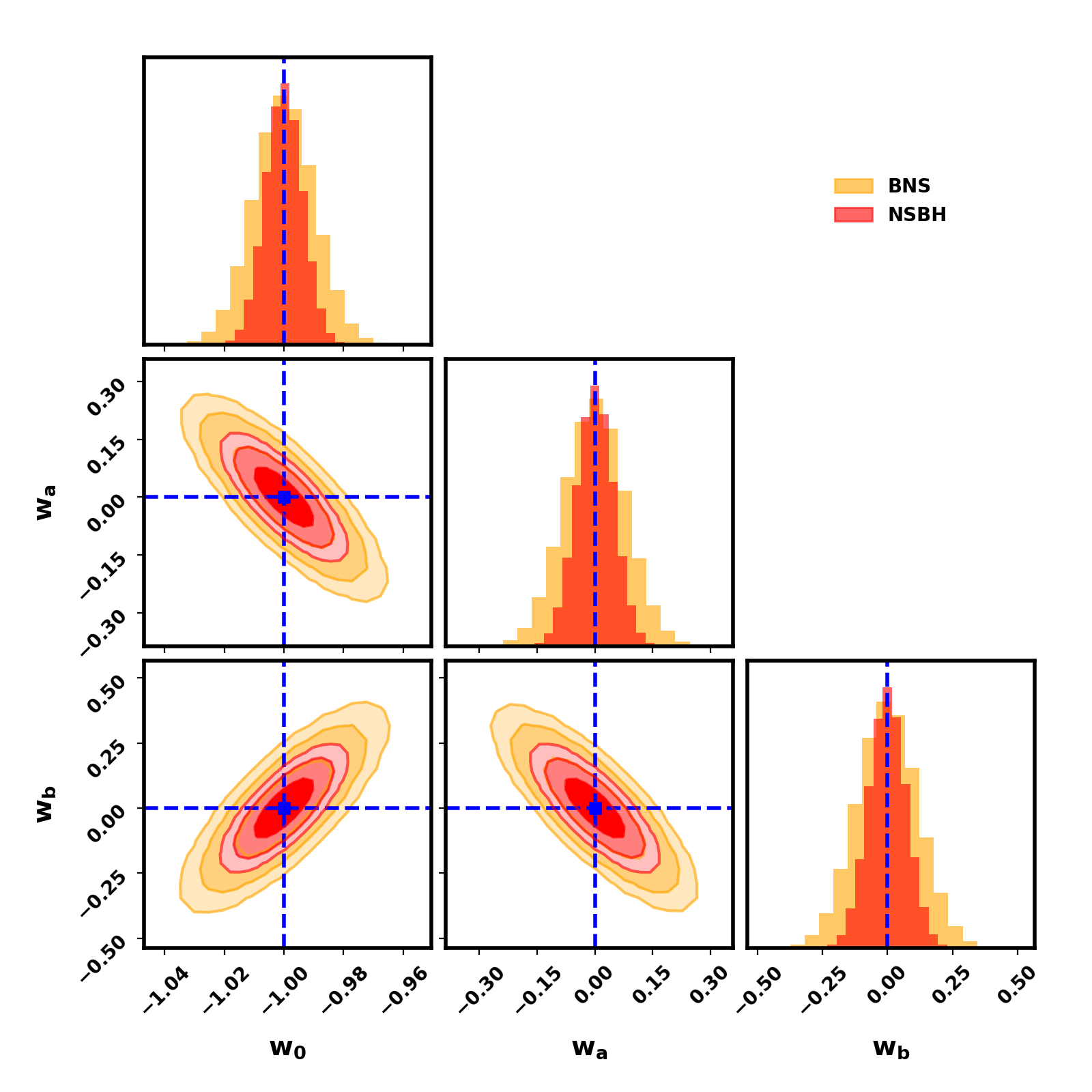}
    \caption{The figure presents the combined posterior distributions of the dark energy equation of state parameters $w_0$, $w_a$, and $w_b$,derived from all detected BNS and NSBH GW events. A total of 75,832 detected BNS and 152,054 detected NSBH events, distributed as shown in Figure \ref{fig:EventCount}, are used in this analysis.}
    \label{fig:DEFullBrightsirens}
\end{figure}

\begin{figure}[ht]
    \centering
    \includegraphics[height=7.65cm, width=10cm]{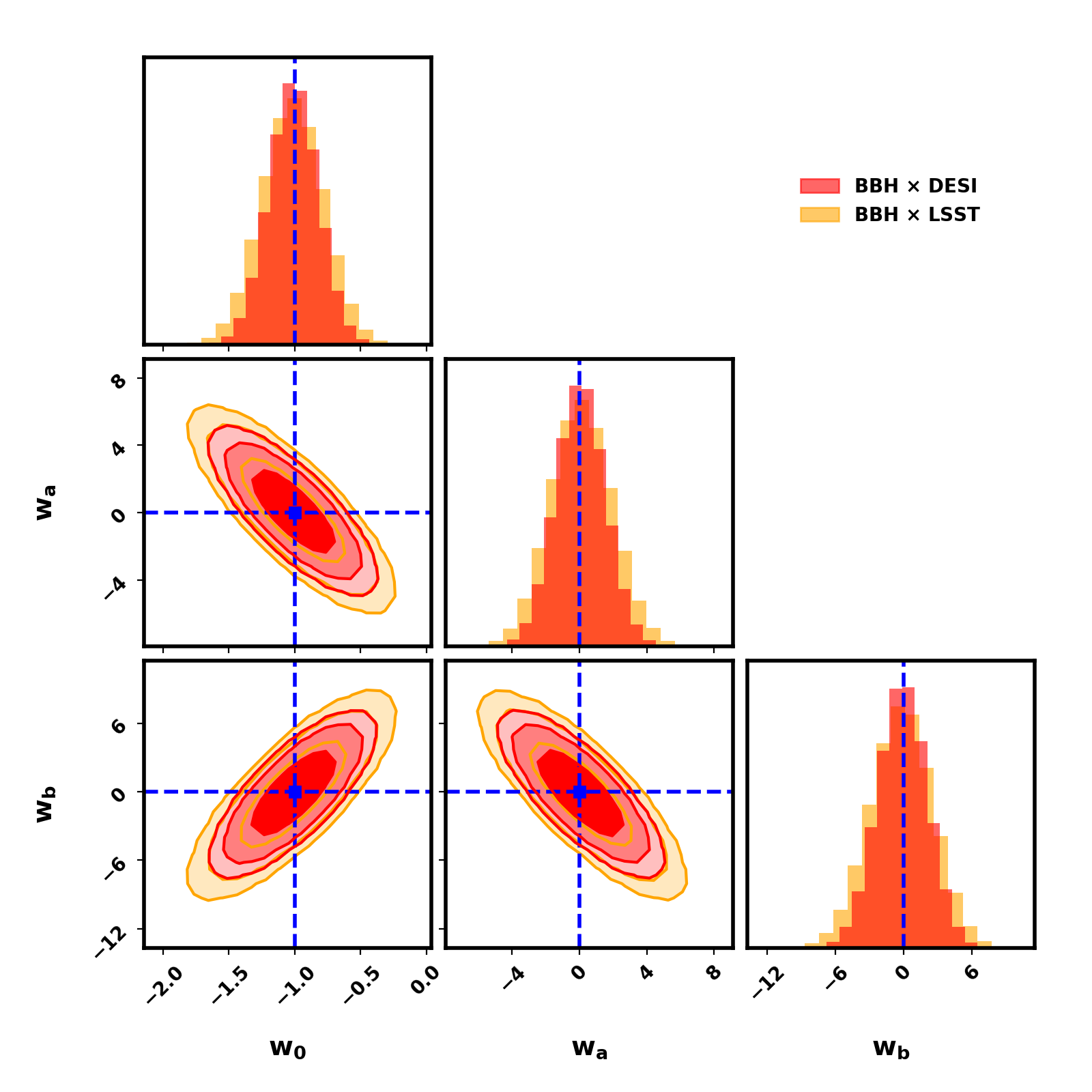}
    \caption{The figure presents the combined posterior distributions of the dark energy equation of state parameters, $w_0$, $w_a$, and $w_b$, derived from all detected BBH sources cross-correlated with the DESI and LSST galaxy surveys. It highlights the constraints on these parameters, demonstrating the significant role of galaxy surveys in improving the measurement of $w(z)$ for dark sirens. A total of 180221 detected BBH events, distributed as shown in figure \ref{fig:EventCount}, are used in this analysis.}
    \label{fig:DEFullDarksirens}
\end{figure}

Figure \ref{fig:DEFullBrightsirens} shows the reconstructed EoS parameters, \( w_0 \), \( w_a \), and \( w_b \), for all GW events across redshifts, derived from bright siren observations under the assumption of a fixed \( H_0 \). The bright siren analysis includes two cases: one for BNS events and another for NSBH events. Similarly, figure \ref{fig:DEFullDarksirens} presents the reconstructed \( w_0 \), \( w_a \), and \( w_b \) parameters for all GW events across redshifts, based on dark siren observations. For the dark sirens, two cases are considered: BBH events cross-correlated with the DESI galaxy catalog, and BBH events cross-correlated with the LSST galaxy catalog.

\begin{figure}[ht]
    \centering
    \begin{minipage}[b]{0.49\textwidth}
        \centering
        \includegraphics[height=7.65cm, width=7.5cm]{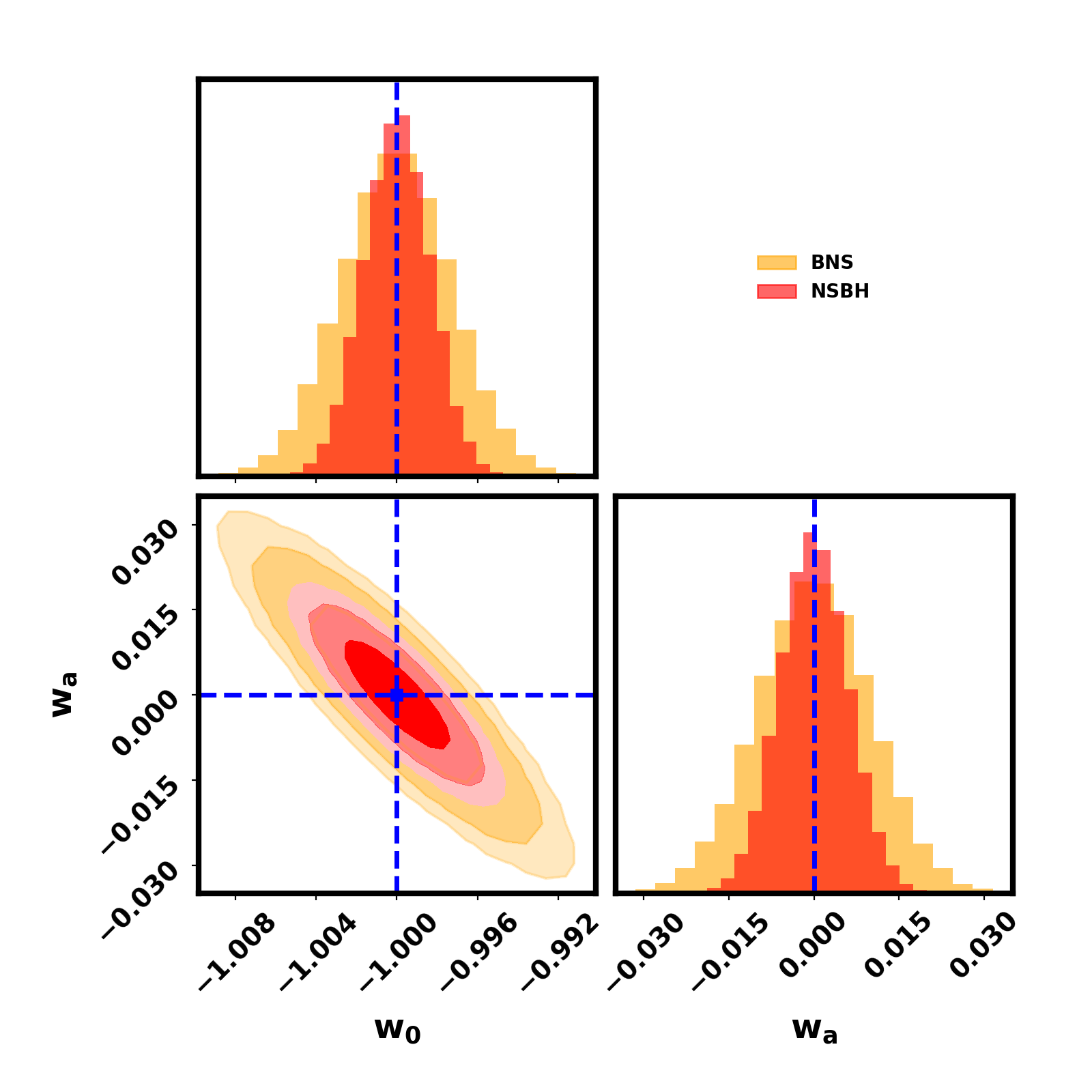}
        \label{fig:DEFullBrightsirensw0wa}
    \end{minipage}
    \begin{minipage}[b]{0.49\textwidth}
        \centering
        \includegraphics[height=7.65cm, width=7.5cm]{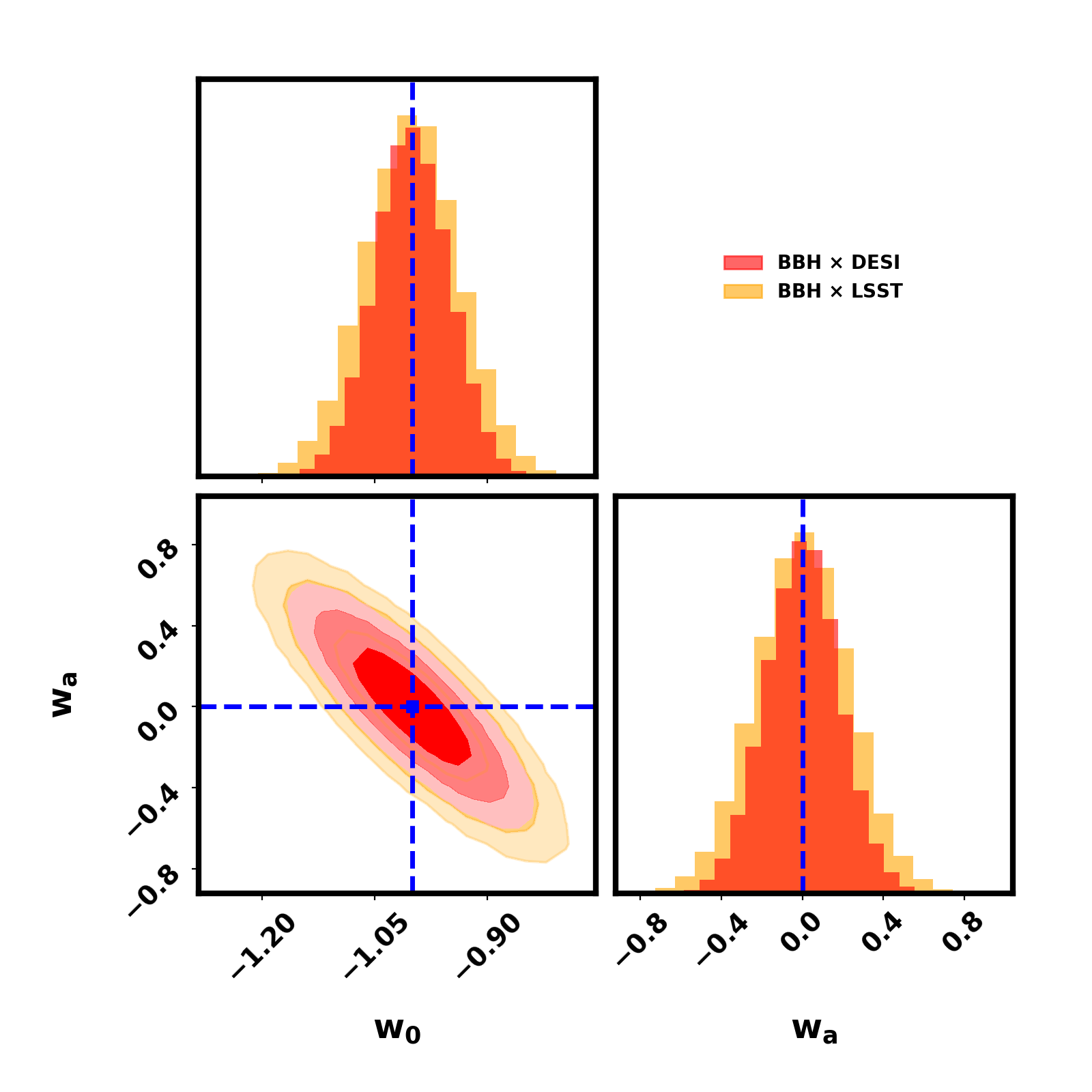}
        \label{fig:DEFullDarksirensw0wa}
    \end{minipage}
    \caption{Combined posterior distributions of the dark energy equation of state parameters, $w_0$ and $w_a$, with $w_b$ held constant at 0, derived from all detected bright and dark sirens. The left panel shows the bright sirens case, highlighting both BNS (binary neutron star) and NSBH (neutron star-black hole) systems. The right panel displays the dark sirens results, showing cross-correlations with the DESI (BBH $\times$ DESI) and LSST (BBH $\times$ LSST) galaxy surveys. A total of 180221 detected BBH, 152054 detected NSBH and 75832 detected BNS events, distributed as shown in figure \ref{fig:EventCount}, are used in this analysis.}
    \label{fig:w0waplot}
\end{figure}

To focus specifically on the CPL parameterization, figure \ref{fig:w0waplot} illustrates the constraints in the \( w_0 \)-\( w_a \) plane under the assumptions of a fixed \( H_0 = 67.4 \, \mathrm{km \, s^{-1} \, Mpc^{-1}} \) and a fixed \( w_b = 0 \). This figure includes results for both bright and dark sirens. For bright sirens, constraints are shown separately for BNS and NSBH events, while for dark sirens, the constraints are illustrated for BBH events cross-correlated with the DESI galaxy catalog and the LSST galaxy catalog. These plots collectively provide a detailed view of how fixing \( H_0 \) influences the reconstructed dark energy EoS parameters. By including all GW events across redshift ranges, the analysis offers a comprehensive perspective on the evolution of dark energy dynamics and highlights the differences in constraints obtained from various types of GW sources and their associated galaxy catalogs.

\bibliographystyle{JHEP.bst}
\bibliography{references} 
\end{document}